%% file: HybDis.tex
\documentclass[useAMS, usenatbib]{mn2e}
\usepackage{amsmath}
\usepackage{amssymb}
\usepackage[dvips]{graphicx,color}
\usepackage{txfonts}
\usepackage[svgnames]{xcolor}
\usepackage{soul}
\usepackage{tikz}
\usetikzlibrary{arrows}
\usetikzlibrary{arrows.meta}

\def\aap{A\&A}
\def\mnras{MNRAS}
\def\apjl{ApJL}
\def\apj{ApJ}
\def\aj{AJ}
\def\sovast{SvA}
\def\araa{ARA\&A}

\def\simless{\mathbin{\lower 3pt\hbox
    {$\rlap{\raise 5pt\hbox{$\char'074$}}\mathchar"7218$}}
}   
\def\simmore{\mathbin{\lower 3pt\hbox
    {$\rlap{\raise 5pt\hbox{$\char'076$}}\mathchar"7218$}}
}   

\def\difd{\mathrm{d}}
\def\mbs{mag\-ne\-to\-brems\-strah\-lung}

\def\nth{{\rm nth}}
\def\mel{m_{\rm e}}
\def\mpr{m_{\rm p}}
\def\acc{{\rm acc}}
\def\inj{{\rm inj}}
\def\BS{B_{\rm S}}
\def\Thetae{\Theta_{\rm e}}
\def\zetae{\zeta_{\rm e}}
\def\sigmaL{\sigma_{\rm L}}
\def\sigmaR{\sigma_{\rm R}}
\def\nchi{{\cal X}}

\def\gminth{\gamma_{\rm min}^{\rm th}}
\def\gmaxth{\gamma_{\rm max}^{\rm th}}
\def\gminnth{\gamma_{\rm min}^{\rm nth}}
\def\gmaxnth{\gamma_{\rm max}^{\rm nth}}

\newcommand{\LL}{\ensuremath{\boldsymbol{\mathcal{L}}}}

\newcommand{\dsfrac}[2]{\displaystyle{\frac{#1}{#2}}}

\newcommand{\unit}[1]{\,\mathrm{#1}}

\newcommand{\modelM}[4]{{\bf M}-{\bf G}{#1}-{\bf D}{#2}-{\bf Z}{#3}-{\LL}{#4}}
\newcommand{\modelW}[4]{{\bf W}-{\bf G}{#1}-{\bf D}{#2}-{\bf Z}{#3}-{\LL}{#4}}

\title[]{On the influence of a Hybrid Thermal-Non thermal distribution in
  the Internal Shocks model for blazars}

\author[J. M. Rueda-Becerril, P. Mimica and M.A. Aloy]
{J. M. Rueda-Becerril$^{1}$\thanks{E-mail:jesus.rueda@uv.es},
  P. Mimica$^{1}$, and M.A. Aloy$^{1}$\\ $^{1}$Departamento de
  Astronom\'{\i}a y Astrof\'{\i}sica, Universidad de Valencia, 46100,
  Burjassot, Spain}

\begin{document}

\maketitle

\label{firstpage}

\begin{abstract}
  Internal shocks occurring in blazars may accelerate both thermal and
  non-thermal electrons. In this paper we examine the consequences that
  such a hybrid (thermal/non-thermal) EED has on the spectrum of
  blazars. Since the thermal component of the EED may extend to very low
  energies. We replace the standard synchrotron process by the more general
  magneto-bremsstrahlung (MBS). Significant differences in the energy flux
  appear at low radio frequencies when considering MBS instead of the
  standard synchrotron emission. A drop in the spectrum appears in the all
  the radio band and a prominent valley between the infrared and soft
  X-rays bands when a hybrid EED is considered, instead of a power-law
  EED. In the $\gamma$-ray band an EED of mostly thermal particles displays
  significant differences with respect to the one dominated by non-thermal
  particles. A thermally-dominated EED produces a synchrotron self-Compton
  (SSC) peak extending only up to a few MeV, and the valley separating the
  MBS and the SSC peaks is much deeper than if the EED is dominated by
  non-thermal particles. The combination of these effects modifies the
  Compton dominance of a blazar, suggesting that the vertical scatter in
  the distribution of FSRQs and BL Lac objects in the peak synchrotron
  frequency - Compton dominance parameter space could be attributed to
  different proportions of thermal/non-thermal particles in the EED of
  blazars. Finally, the temperature of the electrons in the shocked plasma
  is shown to be a degenerated quantity for different magentizations of the
  ejected material.
\end{abstract}


\section{Introduction}
\label{sec:introduction}

In this work we study the emission mechanisms in blazars, a subclass of
radio-loud active galactic nuclei (AGN) in which a relativistic jet is
propagating in the direction very close to the line of sight towards us
\citep[e.g.,][]{Urry:1995aa}. An important observed component of the blazar
radiation is produced by the non-thermal emission from the relativistic jet
they are assumed to host. Its spectrum shows two broad peaks. The first one
is located between radio and X-rays and the second one between X-rays and
$\gamma$-rays \citep[e.g.,][]{Fossati:1998ay}. Depending on the peak
frequencies and the strength of the emission lines blazars can be further
subdivided into BL Lac objects and flat spectrum radio-quasars
\citep[FSRQS; e.g.][]{Giommi:2012aa}. There is a broad consensus that the
low frequency peak is due to the synchrotron emission from the relativistic
electrons gyrating in a magnetic field. As for the high frequency peak,
currently there are two contending models. In the leptonic model the
high-energy emission is produced by the relativistic electrons that
inverse-Compton upscatter both the external low-frequency photons (external
inverse-Compton; EIC), as well as the synchrotron photons produced in the
jet (synchrotron self-Compton; SSC). In the hadronic model there are
relativistic protons in the jet that, in the presence of very strong
magnetic fields, are able to produce the high energy emission both directly
(via proton-synchrotron radiation), as well as via electromagnetic cascades
\citep[see e.g.,][and references therein for a detailed discussion of both
models]{Boettcher:2010aa}. In this work we limit our discussion to the
leptonic model.

The blazar emitted radiation results from the dissipation of the jet
kinetic and Poynting flux. In our work we consider the internal shocks (IS)
model, in which the aforementioned dissipation is produced by the collision
of cold and dense blobs ('shells') within the jet
\citep[e.g.,][]{Rees:1994ca,Spada:2001do,Mimica:2004ay}. Each shell
collision can produce IS that accelerate electrons that are ultimately
responsible for the observed emission.

In the previous papers on this topic we investigated the influence of the
magnetic field on the IS dynamics \citep{Mimica:2010aa} and emission
\citep[][hereafter, the latter two papers will be referred as MA12 and
RMA14, respectively]{Mimica:2007aa,Mimica:2012aa,Rueda:2014mn}. In this
paper we shift our focus to the influence of the properties of the electron
energy distribution (EED) on the observed emission. \citet{Giannios:2009mn}
proposed a mixed Maxwellian/non-thermal EED ('hybrid distribution' or HD
hereafter) as an explanation of some of the features of the gamma-ray burst
prompt and afterglow emission. In this paper we introduce a HD into our
numerical code and study how it affects the blazar light curves and
spectra.

Since the HD thermal component extends to subrelativistic electron
energies, we need to reconsider the emission mechanism (synchrotron) we
employed in previous works. The radiation from charged particles traversing
a magnetic field is known as \mbs\ (MBS). Depending on the speed $\beta c$
of the particles, this radiation is categorized into cyclotron radiation if
$(\beta \ll 1)$ and synchrotron radiation $(\beta \sim 1)$. Both regimes
have been studied broadly and accurate analytical expressions for each have
been developed
\citep[e.g.,][]{Ginzburg:1965gc,Rybicki:1979,Pacholczyk:1970}. However, the
cyclo-synchrotron radiation, i.e., the transrelativistic regime, has no
simple analytic description. Therefore, here we implement a
cyclo-synchrotron (MBS) emission model in our code, to be able to
accurately deal with the emission at all energies of the EED.

In the next section we briefly summarize the dynamics of shell collisions
and the resulting IS. In Secs.~\ref{sec:HD} and~\ref{sec:CySyn} we explain
how the HD and MBS are included in our numerical models. The spectral
differences between the standard synchrotron and MBS, and between a HD and
a power-law EED are presented in Sec.~\ref{sec:cysyn_vs_syn}. In
Sec.~\ref{sec:results} we describe the results from the parameter study of
our model. In Sec.~\ref{sec:temp-vs-mag} we present the electrons
temperature behavior in different IS scenarios. In Sec.~\ref{sec:discuss}
we discuss our results and present our conclusions.

\section{Shell dynamics and emission in the internal shock model}
\label{sec:IS}

We model the shell dynamics and the shock properties in blazar jets as in
MA12. Assuming a cylindrical outflow and neglecting the jet lateral
expansion \citep[it plays a negligible role in blazar jets, see
e.g.,][]{Mimica:2004ay} we can simplify the problem of colliding shells to
a one-dimensional interaction of two cylindrical shells with
cross-sectional radius $R$ and thickness $\Delta r$. The slower (right)
shell Lorentz factor is denoted by $\Gamma_{\rm R}$, while the faster
(left) shell moves with $\Gamma_{\rm L} = (1 + \Delta g)\Gamma_{\rm R}$. In
the previous expression $\Delta g$ stands for the relative Lorentz factor
between the interacting shells. We assume that the shells are initially
cold, so that the fluid thermal pressure ($P$) to rest-mass energy density
ratio $\chi := P/\rho c^2 \ll 1$, where $\rho$ is the fluid rest-mass
density. The shell magnetization is controlled by a parameter
$\sigma := B^2/(4\pi \Gamma^2 \rho c^2)$, where $B$ is the strength of the
large-scale magnetic field (measured in the laboratory frame), that in our
model is assumed to be perpendicular to the shell propagation
direction. Note that the decay of poloidal fields (i.e., parallel to the
shell propagation direction) with distance to the blazar central engine
will be faster than that of toroidal fields (perpendicular to the shell
propagation direction). Certainly, the rate at which the magnetic field
strength may vary with the distance from the blazar central engine depends
on the geometry adopted by the jet. If the jet undergoes a conical
expansion, a decaying power-law with the distance to the central engine is
theoretically expected for the poloidal magnetic field \citep[see
e.g.,][]{Blandford:1974re,Konigl:1981bl}. Pure power-law expressions for
the decay of the magnetic field are roughly adequate until distances
$\sim 1\,$pc from the origin \citep[see
e.g.,][]{Krichbaum:2006vl,Beskin:2006cx,McKinney:2006ou,Asada:2012nz,Nakamura:2013yt,Mohan:2015bg}. Furthermore,
any pre-existing magnetic field component perpendicular to the IS will be
amplified by the standard MHD shock compression. Thus, we expect that the
shells shall possess a magnetic field whose dominant component be
perpendicular to the propagation of shell and, hence, our approximation is
justified.

The number density in an unshocked shell is given by (see equation~3 of
MA12):
\begin{equation}
  \label{eq:numdens}
  n_{i} = \dsfrac{\cal L}{\pi R^2 m_p c^3 \left[\Gamma_{i}^2 (1 + \epsilon
      + \chi +  \sigma_{i}) - \Gamma_{i} \right] \sqrt{1 -
      \Gamma_{i}^{-2}}}\ ,
\end{equation}
where $m_p$ and $c$ are the proton mass and the speed of light, $\epsilon$
is the specific internal energy (see equation~2 of MA12), ${\cal L}$ the
kinetic luminosity of the shells and the index $i = {\rm L, R}$ indicates
which shell we are referring to.

Once the number density, the thermal pressure, the magnetization, and the
Lorentz factor of both shells have been determined, we use the exact
Riemann solver of \citet{Romero:2005zr}, suitably modified to account for
arbitrarily large magnetizations by~\cite{Aloy:2008ApJ}, to compute the
evolution of the shell collision. In particular, we calculate the
properties of the shocked shell fluid (shock velocity, compression factor,
magnetic field) which we then use to obtain the synthetic observational
signature (see the following section). Both in MA12 and RMA14 it is assumed
that a non-thermal EED is injected behind each IS (see e.g., Sec. 3 of
MA12), and the code computes the light curve by taking into account the
synchrotron, SSC and (if needed) EIC processes (Sec. 4 of MA12). The main
modifications introduced by this work are in the hybrid EED injection
spectrum and in the replacement of the pure synchrotron by the MBS
emission.

\section{Hybrid distribution}
\label{sec:HD}

Most IS models for blazars assume that the radiation is produced by a
power-law energy distribution of non-thermal electrons accelerated behind
the shock \citep{Spada:2001do, Mimica:2004ay, Bottcher:2010gn}. More
specifically, the number density of non-thermal particles per unit time and
unit Lorentz factor (both quantities measured in the rest frame of the
fluid%
\footnote{\label{foot:frame}
  The fluid rest frame coincides with the frame of reference of the contact
  discontinuity separating the forward and reverse shocks resulting from
  the collision of two shells, since the fluid in the shocked regions moves
  with the same speed as the contact discontinuity. As we only inject
  particles behind the forward and the reverse shocks, proper fluid
  quantities are identical between these two shocks to those measured in
  the contact discontinuity frame. We note that hereafter, different from
  MA12, we will not annotate with a prime thermodynamical quantities
  measured in the contact discontinuity frame.%
}) is
\begin{equation} \label{eq:pwl-dist}
  \frac{dn_{\nth}}{dt \, d\gamma} = Q_{0} \gamma^{-q} H(\gamma; \gminnth,
  \gmaxnth),
\end{equation}
where $q$ is the power-law index, $\gminnth$ and $\gmaxnth$ are lower
and upper cut-offs for the Lorentz factor of the injected electrons,
respectively, and $Q_{0}$ the normalization coefficient. The interval
function is defined as
\begin{equation}
  H(x;a,b) := \left\{
    \begin{array}{ll}
      1, & a \leq x \leq b \\
      0, & \text{elsewhere}
    \end{array}
  \right..
\end{equation}

As in previous works (\citealt{Mimica:2010cc}; MA12), $\gmaxnth$ is
obtained by assuming that the synchrotron cooling time-scale is
proportional to the gyration time-scale,
\begin{equation}
  \label{eq:g2}
  \gmaxnth = {\left( \frac{3 \mel^{2} c^{4}}{4 \pi a_\acc e^{3} \BS}
  \right)}^{1/2},
\end{equation}
where $e$ is the electron charge, $\mel$ is the electron mass, $\BS$ is the
total magnetic field in the shock and $a_\acc \geq 1$ is the acceleration
efficiency parameter \citep{Bottcher:2010gn,Joshi:2011bp}.

As in MA12 and RMA14, we assume that there exists a
stochastic magnetic field, $B_{\rm{S, st}}$, which is created by the shocks
produced due to the collision of the shells. By definition its strength is
a fraction $\epsilon_{\rm B}$ of the internal energy density of the shocked
shell $u_{\rm S}$ (obtained, in our case, by the exact Riemann solver):
\begin{equation}
  B_{\rm S,st} = \sqrt{8\pi\epsilon_{\rm B}u_{\rm S}}.
  \label{eq:BSst}
\end{equation}
Since we allow for arbitrarily magnetized shells, there is also a
macroscopic magnetic field component, $B_{\rm S,mac}$, which is a
direct output of the exact Riemann solver. The total magnetic field is
then $B:=\sqrt{B^2_{\rm S,st} + B^2_{\rm S,mac}}$.

The motivation for a HD comes from recent PIC simulations of weakly
magnetized relativistic shocks \citep[e.g.,][]{Sironi:2013hf}. These
simulations find that the energy distribution of particles follows a
thermal distribution plus a high energy power-law tail. To describe the
energy distribution of relativistic thermal particles we use the normalized
Maxwell-J\"{u}ttner distribution function
\citep[][p. 394]{Chandrasekhar:1939} so that the number density of thermal
particles per unit time and unit Lorentz factor (both quantities measured
in the rest frame of the fluid) reads
\begin{equation}
  \label{eq:max-jutt}
  \frac{dn_{\rm th}}{dt \, d\gamma} = Q_{\rm th} \frac{\gamma^{2}
    \beta}{\Thetae K_{2}(1 / \Thetae)} e^{-\gamma / \Thetae},
\end{equation}
where $Q_{\rm th}$ is the thermal normalization factor in units of the
number density per unit of proper time, $\gamma$ is the Lorentz factor of
the electrons, $\beta := {(1 - \gamma^{-2})}^{-1/2}$ their velocity,
$\Thetae := k_{\rm B} T / \mel c^{2}$ is the dimensionless electron
temperature, $k_{\rm B}$ is the Boltzmann constant and $K_{2}(x)$ is the
modified Bessel function of second kind. Though the Maxwell-J\"utner
distribution is valid for any Lorentz factor $\gamma\in [1,\infty ]$, for
numerical purposes we limit the previous interval to
$[\gminth,\gmaxth]$. We typically employ
$\gminth = \gamma(\beta = 0.01) \simeq 1.00005$ and $\gmaxth\sim 10^3$.
\citet{Giannios:2009mn} proposed an approximation to a HD (in the GRB
context) consisting of a thermal distribution below a threshold Lorentz
factor and a power-law tail above it. The value of the threshold and the
number of particles in each part is determined by a parameter: the
proportion of non-thermal particles. A similar approach has been used
before by \citet{Zdziarski:1990jk} and \citet{Li:1996pf}, splitting the
distribution at the mean Lorentz factor of the Maxwell-J\"{u}ttner
distribution,
\begin{equation} \label{eq:gamma-th}
  \langle{\gamma}  \rangle = 3 \Thetae + \frac{K_{1}(1 /  \Thetae)}{K_{2}(1
    / \Thetae)}.
\end{equation}

In the standard IS model a fraction $\epsilon_{\rm e}$ of the energy
dissipated at the shock accelerates the electrons into a pure power-law
distribution. In our study we avoid both finding a break Lorentz factor and
estimating the value of $\epsilon_{\rm e}$. Instead we compute the
normalization coefficients of each component by assuming that \emph{all
  thermal energy dissipated at the shock} is used to accelerate
particles. A fraction $\zetae$ of the energy goes into a non-thermal
distribution (the rest going into the thermal part) i.e.,
\begin{equation}
  \label{eq:nthermal-ener-inj}
  \zetae \frac{dE_{\inj}}{dt} = \mel c^{2} V_\acc Q_{0} P(q - 1;  \gminnth,
  \gmaxnth),
\end{equation}
where $V_\acc = \pi R^{2} \Delta r_\acc$ is the volume where the
acceleration takes place (see Sec. 3.2 of MA 12 for more details), $R$ the
cross-sectional radius of the cylindrical shells (which we assume for
simplicity that have the same diameter as the relativistic jet in which
they move), and $Q_{0}$ is the non-thermal normalization factor in units of
number density per unit of time. Equation~\eqref{eq:nthermal-ener-inj} is
obtained by integrating equation~\eqref{eq:pwl-dist} multiplied by
$\gamma \mel c^{2}$ in the interval $[\gminnth, \gmaxnth]$. The function
$P$ is defined as
\begin{equation} \label{eq:pwl-integ}
  P(s; a, b) := \int_{a}^{b} dx\, x^{-s}.
\end{equation}
In a similar way, the fraction of energy injected into the thermal part is
\begin{equation} \label{eq:thermal-ener-inj}
  (1 - \zetae) \frac{dE_{\inj}}{dt} = \mel c^{2} V_\acc  Q_{\rm th}
  \langle{\gamma} \rangle.
\end{equation}

Analogously to the injected energy density, the total number density of
injected particles per unit of proper time is
\begin{equation} \label{eq:tot-num-dens}
  \frac{dn_\inj}{dt} = Q_{\rm th} + Q_{0} P(q; \gminnth, \gmaxnth).
\end{equation}

In analogy to equations (10) and (14) in MA12, the total energy and number
of particles injection rates into the acceleration region are
\begin{align}
  \frac{dE_\inj}{dt} = & \pi R^{2} u_{\rm S} \beta_{\rm S, CD} c,
                         \label{eq:tot-ener-inj} \\
  \frac{dN_\inj}{dt} = & \pi R^{2} n_{i} \Gamma_{i,{\rm CD}}\,
                         \beta_{\rm S, CD} c,
                         \label{eq:tot-num-inj}
\end{align}
where $u_{\rm S}$ is the internal energy density of the shocked shell,
$n_{i}$ is the number density in the shells given by equation
\eqref{eq:numdens}, $\beta_{\rm S, CD}$ is the speed of the shock (see
equation~(5) in MA12) and $\Gamma_{i,{\rm CD}}$ is the bulk Lorentz factor
of each of the shells measured in the contact discontinuity (CD) frame (see
footnote 1).

Assuming that the partition of the number of injected particles is the
same as that of the injected energy we set the following relations
for the normalization coefficients in
equation\,\eqref{eq:tot-num-dens}
\begin{align}
  Q_{0} P(q; \gminnth, \gmaxnth)
  & := \zetae \frac{dn_\inj}{dt} \label{eq:Q0}
  \\
  Q_{th} & := (1 - \zetae) \frac{dn_\inj}{dt} \label{eq:Qth}
\end{align}
From equations~\eqref{eq:Q0} and~\eqref{eq:Qth} we find that
\begin{equation}
  \label{eq:Q0-2}
  Q_{0} = \frac{\zetae Q_{th}}{(1 - \zetae) P(q; \gminnth,
    \gmaxnth)}.
\end{equation}
Finally, from equations~\eqref{eq:nthermal-ener-inj},
\eqref{eq:thermal-ener-inj} and~\eqref{eq:Q0-2} we get the following
expression:
\begin{equation}
  \label{eq:g1}
  P(q - 1; \gminnth, \gmaxnth) =  \langle{\gamma}  \rangle P(q;
  \gminnth, \gmaxnth),
\end{equation}
from which we compute the lower cut-off of the non-thermal distribution
$\gminnth$ using an iterative procedure. For numerical reasons, we do not
allow $\gminnth$ to be smaller than $\gminth$.

Finally, we define the global bounds bracketing both the thermal and
non-thermal EED by
\begin{equation}
\gamma_1=\min{(\gminnth,\gminth)}\quad \text{and}\quad
\gamma_M=\max{(\gmaxnth,\gmaxth)}.
\label{eq:g1_gM}
\end{equation}

\section{Cyclo-synchrotron emission}
\label{sec:CySyn}

Including a thermal distribution of particles implies that low-energy
electrons will also contribute to the emissivity. Here we develop a
formalism that covers the cyclo-synchrotron or MBS emission of both
non-relativistic and relativistic electrons.

For an isotropic distribution of electrons $n(\gamma)$ the emissivity takes
the form \citep[][]{Rybicki:1979}
\begin{equation} \label{eq:emiss-coef-and-power}
  j_{\nu} = \frac{1}{4\pi} \int_{1}^{\infty} d\gamma\, n(\gamma)
  P_{\nu}(\gamma)
\end{equation}
where $P_{\nu}(\gamma)$ is the radiated power of an electron having a
Lorentz factor $\gamma$ and the factor $1 / 4 \pi$ comes from the angular
normalization of the isotropic particle distribution function. These
electrons will spiral around the magnetic field lines, moving with a pitch
angle $\alpha$. The radiated power $P_\nu(\gamma)$, in the comoving frame
(see footnote 1) is
\begin{equation} \label{eq:pow-single-emiss-coeff}
  P_{\nu}(\gamma) = \int_{0}^{2\pi} \int_{-1}^{1} d\phi_{\alpha}\,
  d\mu_{\alpha} \int_{0}^{2\pi} \int_{-1}^{1} d\phi\, d\mu \,
  \eta_{\nu}(\gamma, \vartheta, \alpha),
\end{equation}
where $\phi_{\alpha}$ is the azimuthal pitch angle, $\vartheta$ is the
emission angle (the angle between the emitted photon and the magnetic
field), $\phi$ the azimuthal emission angle, $\mu = \cos\vartheta$,
$\mu_{\alpha} = \cos\alpha$ and the function $\eta_\nu$ is \citep[see
e.g.,][]{Bekefi:1966,Oster:1961he,Melrose:1991},
\begin{eqnarray}
  \eta_{\nu}(\gamma, \mu, \mu_{\alpha})
  & = & \frac{2 \pi e^{2} \nu^{2}}{c}\sum_{m=1}^{\infty} \delta(y_{m})
        \left[ \frac{{(\mu - \beta \mu_ {\alpha})}^{2}}{1 - \mu^{2}}
        J^{2}_{m}(z) \right. + \nonumber
  \\
  & + & \left. \beta^{2} (1 - \mu_{\alpha}^{2}) J'^{2}_{m}(z)
        \right],  \label{eq:single-emiss-coeff}
\end{eqnarray}
where $m$ is an integer index annotating the number of the contributing
harmonic,
\begin{gather}
  y_{m} := \frac{m \nu_{b}}{\gamma} - \nu(1 - \beta \mu_{\alpha}
  \mu), \label{eq:arguments1}
  \\
  z := \frac{\nu \gamma \beta \sqrt{1 - \mu^{2}}
    \sqrt{1 - \mu_{\alpha}^2}}{\nu_{b}}, \label{eq:arguments2}
\end{gather}
$\nu_{b} := e B / 2 \pi \mel c$ is the non-relativistic gyrofrequency,
$J_{m}(x)$ is the Bessel function of the first kind of order $m$, and
$\gamma = 1 / \sqrt{1 - \beta^{2}}$. When the argument in the
$\delta$-function $y_m$ equals zero we met the so-called \emph{resonance
  condition} \citep[also known as the \emph{Doppler condition},
e.g.][]{Leung:2011aj,Melrose:1991}
\begin{equation} \label{eq:reson-cond}
  \frac{m \nu_{b}}{\gamma} - \nu(1 - \beta \mu_{\alpha} \mu) = 0.
\end{equation}
The fulfilment of this condition represents the largest contribution to
the power emitted. For slow electrons $(\beta \ll 1)$, the terms with
small values of $m$ will dominate (manifesting as emission lines), while
for ultrarelativistic ones $(\beta \sim 1)$ the peak of the power radiated
shifts to larger values and the spectrum turns into a continuum. In
Fig.~\ref{fig:chamba-sl-rma} we can observe these features along with the
transrelativistic regime. Since $\eta_{\nu}(\gamma, \vartheta, \alpha)$
depends neither on $\phi_{\alpha}$ nor on $\phi$, the corresponding
integration is straightforward. The final expression for $P_{\nu}(\gamma)$
is then,
\begin{eqnarray}
  P_{\nu}(\gamma)
  & = & \frac{8 \pi^3 e^{2} \nu^{2}}{c} \int_{-1}^{1}
        \int_{-1}^{1} d\mu_{\alpha} d\mu \sum_{m=1}^{\infty}
        \delta(y_{m}) \times \nonumber \\
  & \times & \left[ \frac{{\left( \mu - \beta \mu_{\alpha} \right)}^{2}}{1
             - \mu^2} J^{2}_{m}(z) + \beta^{2} (1 - \mu_{\alpha}^{2})
             J'^{2}_{m}(z) \right]. \label{eq:pow-final-full}
\end{eqnarray}

\subsection{The numerical treatment}
\label{sec:form-behind-chamba}

The numerical evaluation of the MBS emission
(Eq.~\eqref{eq:pow-final-full}) is very challenging because an integral
over an infinite sum of functions $J_{m}$ and their derivatives $J_{m}'$
needs to be performed. Several techniques have been used to compute such
integral. An approximate analytic formula was found by
\citet{Petrosian:1981ij} using the steepest-descent method to achieve good
accuracy in the cyclotron and synchrotron regimes, but the relative errors
in the intermediate regime were between $20$\% and $30$\%. In the subsequent
works an effort has been made to accurately compute the MBS emissivity over
the whole frequency range \citep[for a short review see,
e.g.][]{Leung:2011aj}.

The method we follow consists in first integrating
Eq.~\eqref{eq:pow-final-full} trivially over $\mu_\alpha$, exploiting the
presence of the $\delta$-function. This is the same first step as employed
in~\cite{Leung:2011aj}, but for the Lorentz factors $\gamma$. Then,
from the resonance condition (Eq.~\ref{eq:reson-cond}) we find upper and
lower boundaries for the summation over harmonics. To be more precise, if
$\mu \ne 0$, we solve the resonance condition for $\mu_\alpha$,
\begin{equation}\label{eq:res-cond1}
  \mu_\alpha = \frac{\gamma \nu - m \nu_{b}}{\gamma \nu \beta \mu} =
  \frac{\gamma \nchi - m}{\gamma \nchi \beta \mu},
\end{equation}
where $\nchi := \nu / \nu_{b}$ is the frequency of the emitted photon in
units of the gyrofrequency (also known as \emph{the harmonic number}). The
case $\mu=0$ can be explicitly avoided by performing a numerical
integration of Eq.\,\eqref{eq:pow-single-emiss-coeff} in which none of the
quadrature points falls on zero (see below). Since $|\mu_\alpha| < 1$, the
upper and lower boundaries for the summation in
Eq.~\eqref{eq:single-emiss-coeff} read:
\begin{gather}
  m > \gamma \nchi (1 - \beta \mu) \label{eq:harm-number-lower},
  \\
  m  <  \gamma \nchi (1 + \beta \mu). \label{eq:harm-number-upper}
\end{gather}
Since the values of $m$ must be integer, from
equations~\eqref{eq:harm-number-lower} and~\eqref{eq:harm-number-upper} we
define $m_{+} := \lfloor\gamma \nchi (1 + \beta \mu)\rfloor$ and
$m_{-} := \lceil\gamma \nchi (1 - \beta \mu)\rceil$, obtaining then
from Eq.~\eqref{eq:pow-final-full},
\begin{align}
  P_ {\nu} (\gamma)
  & = \frac{8 \pi^{3} e^{2} \nu_{b} \nchi^{2}}{c}
    \int_ {-1}^ {1} d\mu \, \left( \frac{1}{\nchi \beta |\mu|} \right)
    \times \nonumber \\
  \times & \sum_ {m=m_{-}}^ {m_{+}} \left[ \frac{{\left(\mu - \beta
           \mu_\alpha \right)}^{2}}{1-\mu^2} J^{2}_{m}(z) + \beta^{2}
           (1-\mu_\alpha^2) J'^{2}_{m}(z)
           \right]. \label{eq:powem-no-delta}
\end{align}
where the term in parenthesis before the summation symbol is
$|dy_{m}/d\mu_\alpha|^{-1}$, which comes from the integration of the
$\delta$-function. Note that the value of $\mu_\alpha$ in
Eq.~\eqref{eq:powem-no-delta} must be replaced by the
relation~\eqref{eq:res-cond1}.

Let us now define the following functions:
\begin{align}
  I_{1}(\nchi, \gamma)
  & := \int_{-1}^{1} d\mu \frac{1}{\nchi \beta |\mu|} \times
    \nonumber \\
  \times & \sum_{m=m_{-}}^{m_{+}} \left[ \frac{{\left(\mu - \beta
           \mu_\alpha \right)}^{2}}{1-\mu^2} J^{2}_{m}(z) + \beta^{2}
           (1-\mu_\alpha^2) J'^{2}_{m}(z) \right], \label{eq:I1}
\end{align}
\noindent and
\begin{equation}
  \label{eq:I2}
  \tilde{I}_{2}(\nchi, \gamma_a , \gamma_b) :=
  \int_{\gamma_a}^{\gamma_b} d\gamma\, n(\gamma) \nchi^{2}
  I_{1}(\nchi, \gamma).
\end{equation}
where $\gamma_a$ and $\gamma_b$ are generic input values corresponding to
the upper and lower values of Lorentz factor interval in which the
calculation of equations~\eqref{eq:I1} and~\eqref{eq:I2} will be performed.

In order to compute the emissivity (Eq.~\eqref{eq:emiss-coef-and-power}) we
calculate first $\nchi^{2} I_{1}(\nchi, \gamma)$ and store it in a
two-dimensional array. To minimize the numerical problems caused by a sharp
drop in the power radiated at low Lorentz factors (keeping $\nchi$
constant), a cut-offs array $\{\hat{\gamma}_{\min}\}$ is built (see
Appendix~\ref{sec:interpol-table}). The integration over $\mu$ in
Eq.~\eqref{eq:I1} is performed using a Gauss-Legendre quadrature and
considering the emission to be isotropic. At this stage the evaluation
$\mu = 0$ was avoided by taking an even number of nodal points
(specifically, 120 nodes). To complete the array, we compute the Chebyshev
coefficients in the $\gamma$ direction of $\nchi^2 I_{1}(\nchi, \gamma)$.

\begin{figure}
  \centering
  \includegraphics[width=8cm]{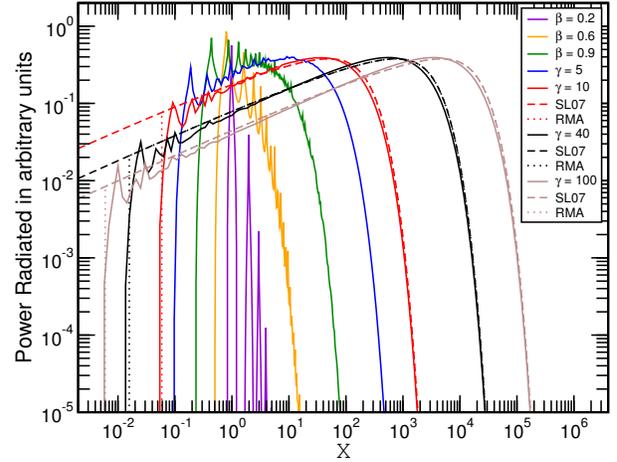}
  \caption{Single electron radiated power as a function of normalized
    frequency computed for different energies (coloured lines) and with
    varying degrees of accuracy.}
\label{fig:chamba-sl-rma}
\end{figure}

The numerical computation of $\nchi^{2} I_{1}(\nchi, \gamma)$ can be made
more efficient taking advantage of the developments by \citet[][hereafter
SL07]{Schlickeiser:2007hp} in order to simplify the computation of the
pitch-angle averaged synchrotron power of an electron having Lorentz factor
$\gamma$, which can be written in the synchrotron limit $(\nchi \gg 1)$ as
\citep{Crusius:1986vu}:
\begin{equation}
  \label{eq:Power-SL07}
  P_\nu^{\rm SL07}(\gamma) = 1.315\times 10^{-28} \nu_b \, x\, CS[x] \:\:
  \text{erg\, sec}^{-1}\text{cm}^{-3},
\end{equation}
where $x:= 2\nchi / (3\gamma^2)$. Comparing the previous expression to
Eq.~\eqref{eq:powem-no-delta} and taking into account
Eq.~\eqref{eq:emiss-coef-and-power} one obtains for sufficiently
relativistic electrons
\begin{equation}
  \label{eq:equiv-relation}
  x\, CS[x] \approx \nchi^{2} I_{1}(\nchi, \gamma).
\end{equation}
The function $CS[x]$ is approximated by (SL07)
\begin{equation}
  \label{eq:CS-approx}
  CS[x] \simeq \frac{x^{-2/3}}{0.869 + x^{1/3} e^{x}},
\end{equation}
which can be computed much faster than the function $I_{1}(\nchi,
\gamma)$. We can use this fact to replace the evaluation of the latter
function by the simpler computation of $CS[x]$ where the appropriate
conditions are satisfied. To determine the region of the parameter space
$(\nchi,\gamma)$ where Eq.~\eqref{eq:CS-approx} holds with sufficient
accuracy we must consider two restrictions. On the one hand, for the first
harmonic, which sets the lower limit where the emissivity is non-zero, we
find that $\nchi_{1}(\gamma) = 1 / \gamma$. On the other hand, the
synchrotron limit (ultrarelativistic limit) happens for $\gamma \gg 1$. For
numerical convenience we take $\gamma_{\rm up}=20$ as a threshold to use
Eq.~\eqref{eq:CS-approx}. For $\gamma > \gamma_{\rm up}$ the evaluation of
$I_{1}$ slows down dramatically since the number of harmonic terms needed
to accurately compute it (Eq.~\eqref{eq:I1}) rapidly increases. To show the
accuracy of the approximations employed in the calculation of $I_{1}$ we
consider the following function (see App.\,\ref{sec:rma-function}):
\begin{equation} \label{eq:rma-func}
  RMA[x] :=
  \begin{cases}
    x\, CS[x] & x >    0.53 / \gamma^3 \\
    0       & \text{otherwise}
  \end{cases}
  ,
\end{equation}
so that, the resulting electron power becomes
\begin{equation}
  P_\nu^{\rm RMA}(\gamma) = 1.315\times 10^{-28} \nu_b \, RMA[x] \:\:
  \text{erg\, sec}^{-1}\text{cm}^{-3},
  \label{eq:Power-RMA}
\end{equation}
In Fig.~\ref{fig:chamba-sl-rma} we show the power radiated by single
electrons with different velocities or, equivalently, Lorentz factors. In
the non-relativistic limit (e.g., for $\beta = 0.2$;
Fig.~\ref{fig:chamba-sl-rma} violet solid line) the spectrum is dominated
by the first few harmonics (first terms in the sum of
Eq.~\eqref{eq:pow-final-full}), which results in a number of discrete peaks
flanked by regions of almost no radiated power. The first harmonic
($m = 1$) peaks at $\nchi\simeq 1$ (a consequence of the resonance
condition, as mentioned above). As the electron velocity increases
($\beta=0.6, 0.9$ and $\gamma=5$; Fig.~\ref{fig:chamba-sl-rma} orange,
green and blue solid lines, respectively) the gaps between the peaks of the
emitted power are progressively filled. In addition, the spectrum broadens
towards ever smaller and larger values of $\nchi$, and an increasing number
of harmonics shows up. At higher Lorentz factors it makes sense to compare
the continuum synchrotron approximation for the electron emitting power
with the MBS calculation. For that we display the cases with
$\gamma=10,\, 40$ and 100 in Fig.~\ref{fig:chamba-sl-rma} with lines
colored in red, black and brown, respectively.  The different line styles
of the latter cases correspond to distinct approximations for the
computation of the MBS power. Solid lines correspond to the numerical
evaluation of Eq.~\eqref{eq:pow-final-full} (the most accurate
result). Dashed lines depict the computation of the synchrotron power as in
SL07 (Eq.~\eqref{eq:Power-SL07}). Dotted lines correspond to the emitted
power calculated according to Eq.~\eqref{eq:Power-RMA}.  The difference
between the three approximations to compute the radiated power decreases as
the Lorentz factor increases. Effectively, for $\gamma > \gamma_{\rm up}$,
both the exact calculation and the approximation given by
$P_\nu^{\rm RMA}(\gamma)$ match rather well. Indeed, the difference becomes
fairly small for $\nchi \gg 1$. The computation of the function $I_{1}$
becomes extremely expensive for large values of $\nchi$, because the number
of harmonics needed to be taken into account for the emitted power to be
computed accurately enough increases dramatically. Thus, in the following
we restrict the more precise numerical evaluation of $P_\nu(\gamma)$
employing $I_{1}$ (Eq.~\eqref{eq:I1}) to cases in which $\nchi \le 100$ and
$\gamma \leq \gamma_{\rm up}$. For $\nchi >100$ and
$\gamma > \gamma_{\rm up}$ we resort to the $RMA$ function
(Eq.~\eqref{eq:rma-func}) to fill in the tabulated values of $I_1$ (see
App.~\ref{sec:interpol-table}). We note that Eq.~\eqref{eq:equiv-relation}
can be generalized to $RMA[x] \approx \nchi^{2} I_{1}(\nchi, \gamma)$, from
which it is obvious that
\begin{equation}
I_{1}(\nchi, \gamma) \approx RMA[x] / \nchi^{2} \qquad \text{if}\quad
  \nchi >100.
\label{eq:RMA-I1}
\end{equation}
We also point to the large quantitative and qualitative effect of the
cut-off in the emitted power resulting from the use of the $RMA[x]$
function (Eq.~\eqref{eq:rma-func}) in the evaluation of
$P_\nu^{\rm RMA}(\gamma)$ (Eq.~\eqref{eq:Power-RMA}). This cut-off is in
contrast to the non-zero emitted power at low frequencies, a characteristic
of the synchrotron (continuum) approximation.

\subsection{Numerical evaluation of the emissivity}
\label{sec:table}

In this section we describe how an interpolation table is built and
afterwards used to compute the emissivity
(Eq.~\eqref{eq:emiss-coef-and-power}) numerically. We discretize the
HD by tessellating it in a large number of Lorentz-factor intervals
whose boundaries we annotate with ${\{ \gamma_{i} \}}_{i=1}^{M}$. Note
that the smallest and largest value of the Lorentz factor tessellation
coincide with the definitions given in Eq.\,\eqref{eq:g1_gM}.  For
numerical convenience and efficiency, in every interval we approximate
the EED by a power-law function (with a power-law index $q_{i}$)
since, for this particular form it is possible to analytically perform
a part of the calculation, which drastically reduces the computational
time. Then we use the new interpolation table to compute the
emissivity at arbitrary frequency as described below.

\subsubsection{The construction of the interpolation table}
\label{sec:construction}

Performing a direct numerical integration of Eq.~\eqref{eq:I2} may lead to
numerical noise in the final result due to the extremely large amplitude
oscillations of the integrand in the limits $\nchi \ll 1$ and
$\gamma \simeq 1$. Therefore, assuming a power-law distribution, we
reformulated $\tilde{I}_{2}$ in the following manner
\begin{align}
  I_{2}(\nchi, q, \gamma_{i}, \gamma_{i+1};
  & \gmaxth) = {\left( \gmaxth\right)}^{1-q} \times \nonumber \\
  & \times \int_{\gamma_{i} / \gmaxth}^{\gamma_{i+1} / \gmaxth} \difd\xi\,
    \xi^{-q} \nchi^{2} I_{1}(\nchi, \xi\gmaxth) \label{eq:I2-reformulated}
\end{align}
where $q$ is the index of the power-law approximation to the EED within the
interval $[\gamma_{i}, \gamma_{i+1}]$ and $\xi := \gamma / \gmaxth$.  When
calculating $\tilde{I}_2$ (Eq.\,\eqref{eq:I2}), an integral with this shape
suggests the definition of
\begin{equation}
  \label{eq:I3}
  I_{3}(\xi, \nchi, q) := \int_{\xi}^{1} \difd\hat{\xi}\, \hat{\xi}^{-q}
  \nchi^{2} I_{1}(\nchi, \hat{\xi} \gmaxth),
\end{equation}
where $\hat{\xi}$ is an ancillary variable. Rewriting $I_{2}$ in terms of
$I_{3}$ we get
\begin{align}
  \displaystyle
  I_{2}(\nchi, q, \gamma_{i}, \gamma_{i+1};
  & \,\gmaxth) = {\left( \gmaxth \right)}^{1-q}\,\times \nonumber \\
  & \times \left[ I_{3}\left( \frac{\gamma_{i}}{\gmaxth}, \nchi, q
    \right) - I_{3}\left( \frac{\gamma_{i+1}}{\gmaxth},
    \nchi, q \right) \right]. \label{eq:I2-of-I3}
\end{align}

We calculate the integral that depends on the three parameters in
Eq.~\eqref{eq:I3} resorting to a standard Romberg quadrature method for
each value of the triplet $(\xi, \nchi, q)$. In the same manner as with
Eq.~\eqref{eq:I1}, a three dimensional array is built for
$I_{3}(\xi, \nchi, q)$ with the Chebyshev coefficients in the $\xi$
direction in order to construct an interpolation table for $I_{2}$
(hereafter {\tt disTable}).

The integral over Lorentz factors was performed for all values of $\nchi$
and $q$ using a Romberg integration routine. Analogously to $\tilde{I}_{1}$
(see App.~\ref{sec:reconstr-radi-power}), the Chebyshev polynomials were
constructed in the $\xi$ direction.

\subsubsection{Computation of emissivity using an interpolation table}
\label{sec:reconstruction}

In terms of $I_2$ (Eq.~\ref{eq:I2-of-I3}), the evaluation of the emissivity
(Eq.~\eqref{eq:emiss-coef-and-power}) in any of the power-law segments in
which the original distribution has been discretized, e.g., extending
between $\gamma_i$ and $\gamma_{i+1}$ and having a power-law index $q_i$,
reads
\begin{equation}
  \label{eq:emiss-coef-from-table}
  j_{\nu, i} = \frac{\pi e^2 \nu_{b}}{2 c} n(\gamma_{i}) \gamma_i^{q_i}
  I_{2} \left(\nchi, q_i, \gamma_i, \gamma_{i+1}; \gmaxth\right).
\end{equation}
Then, the total emissivity from an arbitrary EED can efficiently be
computed by adding up the contributions from all power-law segments
\citep[see e.g. Sec. 4 in][]{Mimica:2009aa}.

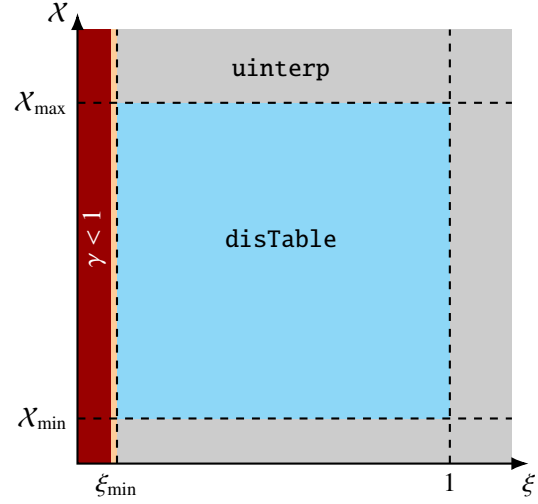
\begin{figure}
  \centering
  \input{Figures/Fig02}
  \caption{Illustration of the different regions of the $\nchi-\xi$ space
    spanned by the distinct approximations employed to compute the values
    of emissivity according to
    Eq.~\eqref{eq:emiss-coef-from-table}. $\nchi_{\min}$ and $\nchi_{\max}$
    are generic values for upper and lower limits of $\nchi$ for the table
    {\tt disTable} and $\xi_{\min} \equiv \gminth / \gmaxth$. For a given
    $q$, a combination of $\xi$ and $\nchi$ in the blue region means that
    {\tt disTable} is employed. The red area corresponds to the physically
    forbidden regime where $\gamma < 1$ and, therefore, there is no MBS
    emission. The thin orange strap corresponds to the area of low speeds
    $1 \leq \gamma < \gamma_1$ excluded from the table.}
\label{fig:tables-scheme}
\end{figure}

The discretization of {\tt disTable} in the $(\nchi, \xi)$-plane is not
uniform. Many more points are explicitly computed in the regime
corresponding to low electron energies and emission frequencies than in the
rest of the table. In this regime harmonics dominate the emissivity and
accurate calculations demand a higher density of tabular points. In the
ultrarelativistic regime the emission is computed also numerically. For
that we resort to the table produced in MA12 (hereafter {\tt uinterp}) that
includes only the synchrotron process computed with relative errors smaller
than $10^{-5}$. Note that in the ultrarelativistic regime the errors made
by not including the contribution of the MBS harmonics are negligible. We
use both tables in order to cover a wider range of frequencies and Lorentz
factors than would be possible if only {\tt disTable} were to be used (due
to the prohibitively expensive calculation for high frequencies and Lorentz
factors). In Fig.~\ref{fig:tables-scheme} we sketch the different regions
of the $\nchi - \xi$ space spanned by our method to assemble a single
(large) table.  Whenever our calculations require the combination of
$\nchi$ and $\xi$ that falls in the blue region, we employ {\tt disTable}
to evaluate the emissivity, otherwise we use {\tt uinterp}. In the
particular case when $\gamma_i < \gmaxth <\gamma_{i+1}$, the emissivity is
computed using both tables as follows:
\begin{align}
  j_{\nu, i} =
  \frac{\pi e^2 \nu_{b}}{2 c} n(\gamma_i) \gamma_i^{q_i}
  \times & \left( I_{2}^{\mathtt{disTable}}(\nchi, q_i, \gamma_i, \gmaxth;
           \gmaxth) + \right. \nonumber \\
         &\left.  I_{2}^{\mathtt{uinterp}}(\nchi, q_i, \gmaxth,
           \gamma_{i+1}; \gmaxth) \right).  \label{eq:jnu-matchup}
\end{align}

\section{Differences between MBS and standard synchrotron spectra}
\label{sec:cysyn_vs_syn}

In this section we show the importance of the introduction of the new MBS
method into our blazar model. We will first show the differences that arise
from using different approximations for the emission process assuming the
same HD with a dominant non-thermal component (Sect.~\ref{sec:fixedHD})
for each test. In the second test we compare the spectra produced by a
non-thermally dominated HD with that of a pure power-law extending towards
$\gamma_1\simeq 1$ (Sect.~\ref{sec:compar-g1sim1}) by computing both MBS
and pure synchrotron emission.

For the evolution of the particles injected at shocks, we assume that the
dominant processes are the synchrotron cooling and the inverse-Compton
scattering off the photons produced by the MBS processes (SSC\footnote{For
  simplicity we keep the abbreviation ``SSC'' to denote the process of
  scattering of the non-thermal emission produced by the local electrons
  off those same electrons, but it should be noted that in our model the
  seed photons for the inverse-Compton scattering are produced by the (more
  general) cyclo-synchrotron emission (Sec.~\ref{sec:CySyn}).}). We note
that, in many cases, SSC cooling may be stronger than synchrotron cooling,
as we shall see in Sect.~\ref{sec:results}.  To compute synthetic
time-dependent multiwavelength spectra and light curves, we include
synchrotron and synchrotron self-Compton emission processes resulting from
the shocked plasma. We further consider that the observer's line of sight
makes an angle $\theta$ with the jet axis. A detailed description of how
the integration of the radiative transfer equation along the line of sight
is performed can be found in Section~4 of MA12.

To avoid repeated writing of the parameter values when referring to
our models, we introduce a naming scheme in which the magnetization is
denoted by the letters {\bf S}, {\bf M} and {\bf W}, referring to the
following families of models:
\begin{itemize}
\item[{\bf W}:] weakly  magnetized, $\sigmaL = 10^{-6}, \sigma_{\rm R} =
  10^{-6}$,
\item[{\bf M}:] moderately magnetized, $\sigmaL = 10^{-2}, \sigma_{\rm R} =
  10^{-2}$, and
\item[ {\bf S}:] strongly magnetized, $\sigmaL = 10^{-1}, \sigma_{\rm R} =
  10^{-1}$.
\end{itemize}
The remaining four parameters ${\cal L}$, $\Gamma_{\rm R}$, $\Delta g$ and
$\zetae$ can take any of the values shown in
Table~\ref{tab:parameters}. When we refer to a particular model we label it
by appending values of each of these parameters to the model letter. For
the parameter $\zetae$ we use {\bf Z}m2, {\bf Z}m1 and {\bf Z}09 to refer
to the values $\zetae = 10^{-2}, 10^{-1}$ and $0.9$,
respectively. Similarly, for the luminosity we write {\LL}1, {\LL}5, and
{\LL}50 to denote the values $10^{47} \unit{erg} \unit{s}^{-1}$,
$5 \times 10^{47} \unit{erg} \unit{s}^{-1}$ and
$5 \times 10^{48} \unit{erg} \unit{s}^{-1}$, respectively. In this
notation, \modelW{10}{1.0}{m1}{5} corresponds to the weakly magnetized
model with $\Gamma_{\rm R}=10$ ({\bf G}10), $\Delta g = 1.0$ ({\bf D}1.0),
$\zetae=0.1$ ({\bf Z}m1) and
${\cal L} = 5 \times 10^{47} \unit{erg} \unit{s}^{-1}$ ({\LL}5).

\subsection{Spectral differences varying the emissivity for a fixed HD}
\label{sec:fixedHD}

\begin{figure}
  \centering
  \includegraphics[width=8.5cm]{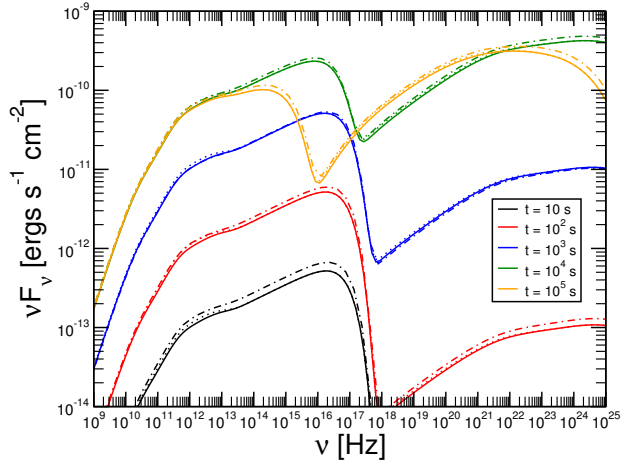}
  \caption{Instantaneous spectra for a model including a HD in which 90\%
    of the particles populate the non-thermal tail of the EED computed
    employing our new MBS numerical method (full lines), using the direct
    numerical integration of $RMA[x]$ function (dotted lines, see
    Eq.~\ref{eq:rma-func}), and using the direct numerical integration of
    the \citet{Crusius:1986vu} function (dot-dashed lines). The dynamical
    model employed corresponds to a collision of weakly magnetized shells.}
\label{fig:instant-spec-ze0p9}
\end{figure}

In Fig.~\ref{fig:instant-spec-ze0p9} we display the instantaneous spectra
of a weakly magnetized model containing a HD where 90\% of the particles
populate the non-thermal tail of the EED (model \modelW{10}{1.0}{09}{1})
taken at $10$, $10^2$, $10^3$, $10^4$ and $10^5$ seconds after the start of
the shell collision. Solid, dotted and dashed lines show the emission
computed using the full MBS method (Sec.~\ref{sec:form-behind-chamba}) and
the direct numerical integration of the analytic approximations $RMA[x]$
(Eq.~\ref{eq:rma-func}) and the numerical integration of the
\citet{Crusius:1986vu} function employed in MA12 and RMA14 (referred
hereafter as the standard synchrotron), respectively. The difference
between the first two and the third is in the presence of a low-frequency
cut-off which causes appreciable differences at early times. The purely
synchrotron emission (dot-dashed lines) always produces an excess of
emission with respect to the other two. This is explained by the fact that
there is always a portion of the EED whose energy is too low for it to be
emitting in the observed frequencies in a more realistic MBS model (see
Fig.~\ref{fig:chamba-sl-rma}). The approximate formula $RMA[x]$ performs
quite well and its spectra mostly overlap the MBS ones, except close to the
first turnover in the spectrum (corresponding to the maximum of the
emission from the lowest-energy electrons). Despite the presence of a
cutoff in $RMA[x]$, it still overestimates the low-frequency emission just
below the first harmonic, which explains the observed slight mismatch.

\subsection{Spectral differences between an HD and a pure power-law EED}
\label{sec:compar-g1sim1}

In the previous section we have seen that the differences between the MBS
emissivity and the pure synchrotron emissivity are relatively mild if we
consider a hybrid, non-thermally dominated EED. To a large extend this
happens because a HD is flanked by a monotonically decaying tail at low
electron energies (which indeed goes to zero as the electron Lorentz factor
approaches 1, see inset of Fig.~\ref{fig:comp-g1sim1}). Here we are
interested in outlining the spectral differences when the lower boundary of
the EED is varied. For that we consider two different EEDs, namely, a
non-thermally dominated HD (corresponding to model \modelW{10}{1.0}{09}{1})
and a pure power-law EED extending to $\gamma_1\simeq 1$. The rest of the
parameters of our model, including the MBS emissivity are fixed. To set up
the pure power-law EED we cannot follow exactly the same procedure as
outlined in Sect.~\ref{sec:HD} because we must fix $\gamma_1$ instead of
obtaining it numerically solving Eq.~\ref{eq:g1}. Furthermore, we employ
the same non-thermal normalization factor $Q_0$ for both the pure power-law
EED and the HD.

In Fig.\,\ref{fig:comp-g1sim1} we show the spectral energy distribution
corresponding to both the HD and pure power-law EED cases. It is evident
that there are substantial differences at frequencies below the GHz range
and in the infrared-to-X-rays band.  On the other hand, the synchrotron
tails above $\sim 10^{13}$\,Hz are almost identical for both
EED. Correspondingly, the cyclo-synchrotron photons there produced are
inverse Compton upscattered forming nearly identical SSC tails above
$\sim 10^{20}$\,Hz.

\subsection{Spectral differences between MBS and pure synchrotron for the
  same power-law distribution}
\label{sec:mbs-vs-syn}

\begin{figure}
  \centering
  \includegraphics[width=\columnwidth]{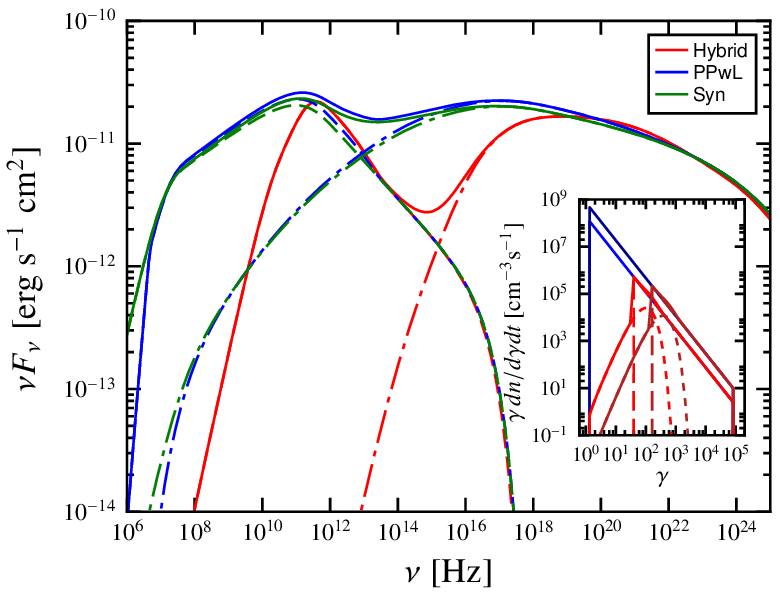}
  \caption{Comparison between the same hybrid model as in
    Fig.~\ref{fig:instant-spec-ze0p9} and a pure power-law distribution
    with $\gminnth \simeq 1$. The red lines correspond to the former model
    while the green and blue lines correspond to simulations with the
    latter distributions using our MBS numerical method and numerical
    integration of \citet{Crusius:1986vu}, respectively. Dashed and dot
    dashed lines show the synchtron and SSC spectral contributions to each
    of the respective models. Inset: the injected EEDs in each shock. Blue
    and dark blue colors correspond to the EED for a pure power-law
    distribution injected at the FS and at the RS, respectively. Red and
    dark red colors correspond to the HD distribution injected at the FS
    and at the RS, respectively.}
\label{fig:comp-g1sim1}
\end{figure}

In the previous section we pointed out how different the SEDs may result
for different distributions. Let us now fix the same injected power-law EED
starting from $\gamma_{1} \approx 1$ and evaluate the emissivities
corresponding to MBS and pure synchrotron processes. In both cases the SSC
is also computed. In Fig.~\ref{fig:comp-g1sim1} we included the averaged
SED from a simulation with the same configuration as the pure power-law EED
model mentioned above but the radiation treatment was numerical standard
synchrotron (green lines). From $10^{10}$\,Hz to $10^{22} \unit{Hz}$ the
MBS spectrum is quite similar to that of a pure synchrotron one, so that
both emission models are observationally indistinguishable in the latter
broad frequecy range. On the other hand, if we look into the MHz band, we
will find what we call the cyclotron break, which is the diminishing of the
emissivity from each electron due to the cut-off that happens at
frequencies below $\nu_{b}$.

\section{Parameter study}
\label{sec:results}

In order to assess the impact of the presence of a hybrid distribution
composed by thermal and non-thermal electrons we have performed a
parametric study varying a number of intrinsic properties of the shells. In
the following subsections we examine the most important results of our
parametric study. In the Tab.~\ref{tab:parameters} we show the values of
the parameters used in the present work. Some of them are fixed in the
following and are shown with a single value in
Tab.~\ref{tab:parameters}. Among such parameters, we find the fraction of
the internal energy density of the shocked shell converted into stochastic
magnetic field energy density, $\epsilon_{\rm B}$, the size of the
acceleration zone, $\Delta_{\rm acc}$, and the number of turns around
magnetic field lines in the acceleration zone that electrons undergo before
they cool down, $a_{\rm acc}$ \citep[see][for further
details]{Mimica:2012aa}. The cross-sectional radius and longitudinal size
of the shells are given by the parameters $R$ and $\Delta r$, respectively.

One of the parameters kept constant in the previous studies is the total
jet luminosity ${\cal L}$, which we now vary. We performed a number of test
calculations to compute the lower and upper limits of ${\cal L}$ that
produce a spectrum qualitatively similar to that of the source \emph{Mrk
  421} \citep{Krawczynski:2013fr}. In the Table~\ref{tab:parameters} we
show the range of variations of this and other parameters.

We perform our parametric scan for the typical redshift value of \emph{Mrk
  421}, namely, $z = 0.031$. The viewing angle is fixed to $5^\circ$ in all
our models. The SEDs in this work were computed by averaging over a time
interval of $10^{7} \unit{s}$.
\begin{table}
  \centering
  \begin{tabular}{cc}
    \hline \hline
    Parameter & value \\
    \hline
    $\Gamma_{\rm R}$ & $2,\ 10,\ 20$\\
    $\Delta g$ & $1.0,\ 2.0,\ 3.0,\ 5.0$\\
    $\sigmaL$ & $10^{-6},\ 10^{-2},\ 10^{-1}$\\
    $\sigmaR$ & $10^{-6},\ 10^{-2},\ 10^{-1}$\\
    $\epsilon_{\rm B}$ & $10^{-3}$ \\
    $\zetae$ & $10^{-2},\ 10^{-1},\ 0.9$ \\
    $q$ & $2.6$ \\
    $\Delta_{\rm acc}$ & $10$ \\
    $a_{\rm acc}$ & $10^{6}$ \\
    ${\cal L}$ & $10^{47},\ 5\times 10^{47},\ $ $5\times 10^{48} \unit{erg}
                 \unit{s}^{-1}$ \\
    $R$ & $3\times 10^{16}$ cm \\
    $\Delta r$ & $6\times 10^{13}$ cm \\
    $z$ & $0.031$ \\
    $\theta$ & $5^\circ$ \\
    \hline
  \end{tabular}
  \caption{Model parameters. $\Gamma_{\rm R}$
    is the Lorentz factor of the slow shell,
    $\Delta g:=\Gamma_{\rm L}/\Gamma_{\rm R} - 1$ ($\Gamma_{\rm L}$ is the
    Lorentz factor of the fast shell), $\sigmaL$ and $\sigma_{\rm R}$ are
    the fast and slow shell magnetizations, $\epsilon_{\rm B}$ is the
    fraction of the internal energy density at shocks that it is assumed to
    be converted into stochastic magnetic field energy density
    \mbox{(Eq.~\eqref{eq:BSst})}, $\zeta_e$ and $q$ are the fraction of
    electrons accelerated into power-law Lorentz factor (or energy)
    distribution and its corresponding power-law index$^{\ref{foot:qval}}$,
    $\Delta_{\rm acc}$ and $a_{\rm acc}$ are the parameters controlling the
    shock acceleration efficiency (see Section~3.2 of MA12 for details),
    ${\cal L}$, $R$ and $\Delta r$ are the jet luminosity, jet radius and
    the initial width of the shells, $z$ is the redshift of the source and
    $\theta$ is the viewing angle. Note that $\Gamma_{\rm R}$, $\Delta g$,
    $\sigmaL$, $\sigma_{\rm R}$ and $\zetae$ can take any of the values
    indicated.}
\label{tab:parameters}
\end{table}
\addtocounter{footnote}{1} \footnotetext{\label{foot:qval} The chosen value
  for $q$ is representative for blazars according to observational
  \citep{Ghisellini:1998hg} and theoretically deduced values
  \citep{Kardashev:1962sv,Bottcher2002pd}. It also agrees with the ones
  used in numerical simulations of blazars made by~\cite{Mimica:2004PhD}
  and~\cite{Zacharias:2010aa}.  }

\subsection{The presence of the non-thermal population}
\label{sec:non-therm-popul}

\begin{figure*}
  \centering
  \includegraphics[width=8.5cm]{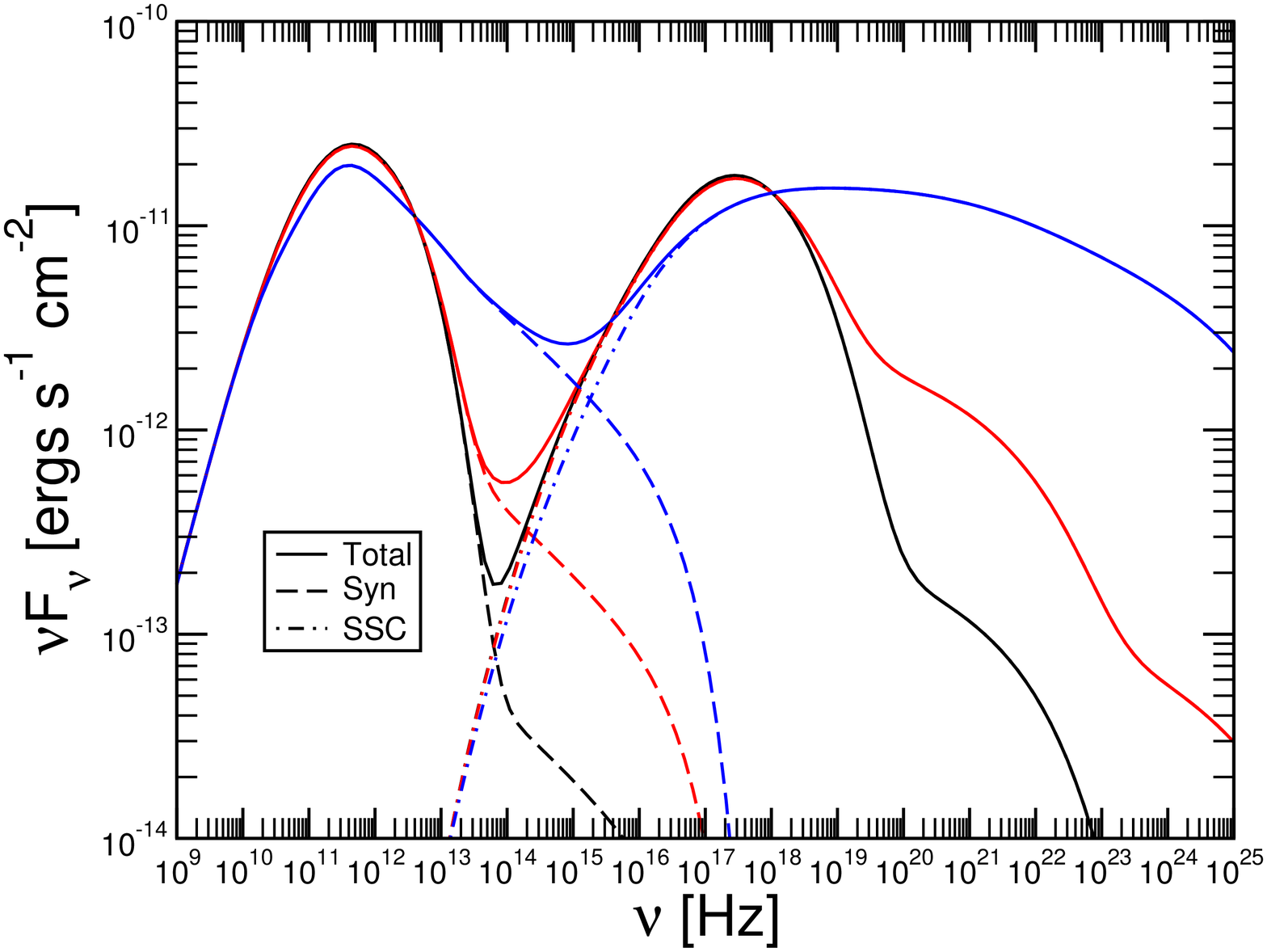}
  \hspace{0.16cm}
  \includegraphics[width=8.5cm]{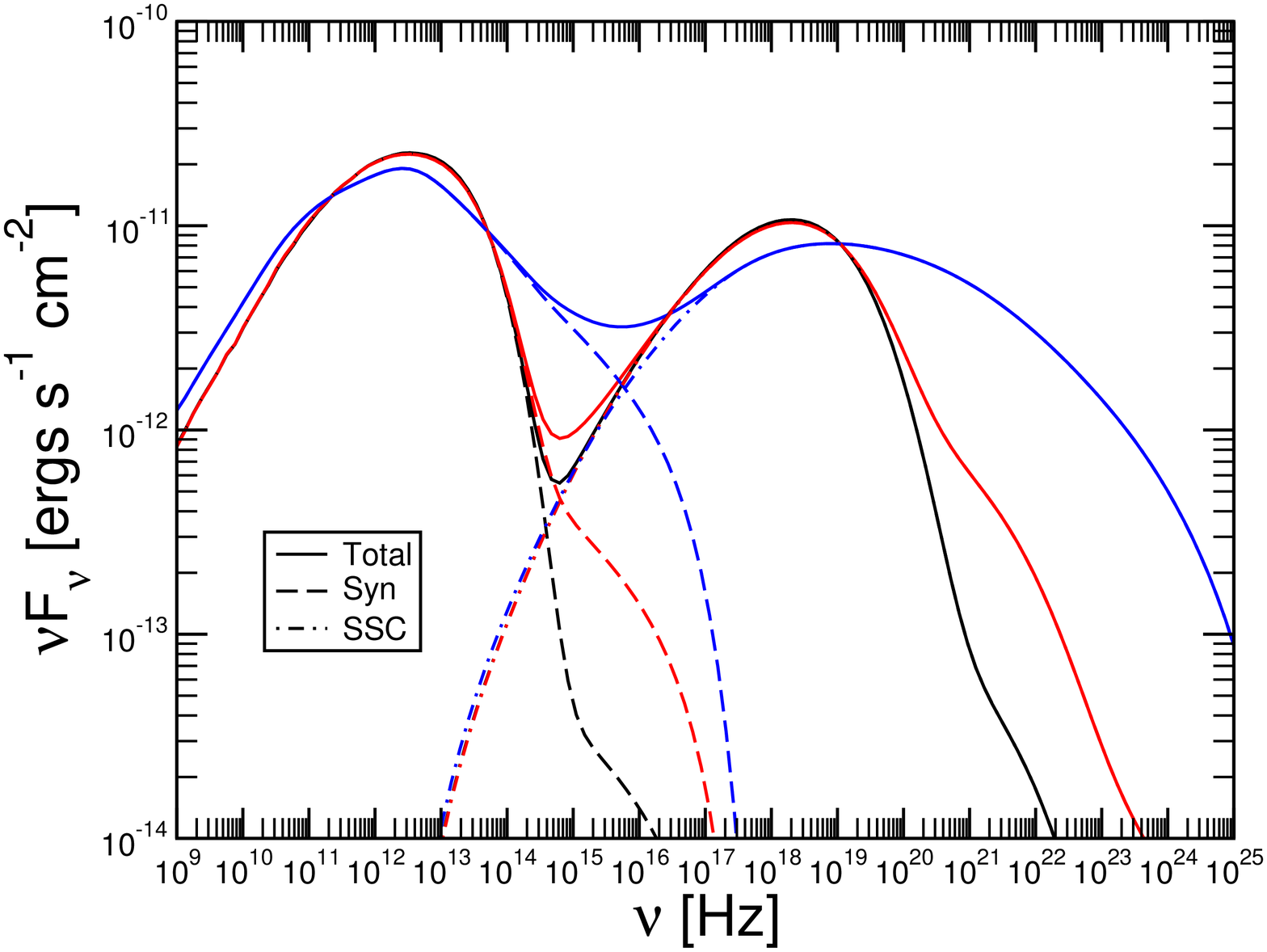}
  \caption{Averaged spectra of the weakly (left panel) and moderately
    (right panel) magnetized models for $\zetae = 0.9, 0.1$ and $0.01$ in
    blue, red and black lines respectively. Dashed lines show the
    synchrotron component while the dot-dashed lines show the SSC
    component.}
\label{fig:nonthermal-pop}
\end{figure*}

The influence of the parameter $\zetae$ on the blazar emission was examined
in \citet{Bottcher:2010gn}, and is an essential model parameter in MA12 and
RMA14 as well (though in the latter two papers it was not varied). In this
section we explore its influence it by studying three different fractions
of non-thermal particles: $\zetae = 0.9, 0.1, 0.01$. In
Fig.~\ref{fig:nonthermal-pop} we show the averaged SEDs of the models with
the aforementioned values of $\zetae$ for the weakly (left panel) and
moderately (right panel) magnetized shells. In both panels we can
appreciate that an EED dominated by non-thermal particles produces a
broader SSC component. The SSC component of a thermally-dominated EED
(\modelW{10}{1.0}{09}{5} and \modelM{10}{1.0}{09}{5}) displays a steeper
synchrotron-SSC valley, and the modelled blazar becomes $\gamma$-rays
quiet. The synchrotron peak frequency $\nu_{\rm syn}$ is only very weakly
dependent on $\zetae$. According to their synchrotron peak frequency these
models resemble low synchrotron peaked blazars (LSP)
\citep{Giommi:2012aa,Giommi:2013fd}.

\subsection{Magnetization}
\label{sec:magnetization}

\begin{figure*}
  \centering
  \includegraphics[width=8.5cm]{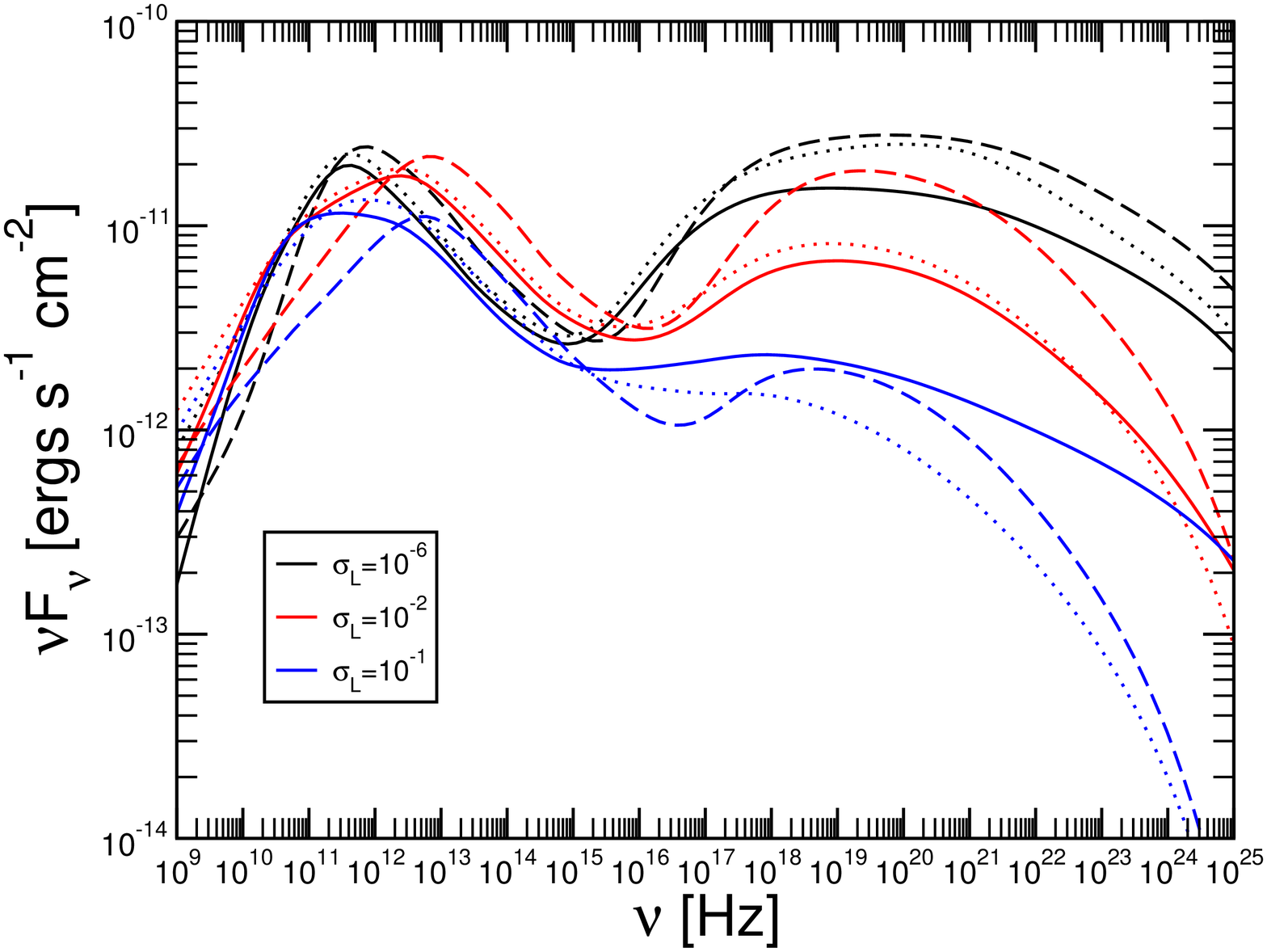}
  \hspace{0.16cm}
  \includegraphics[width=8.5cm]{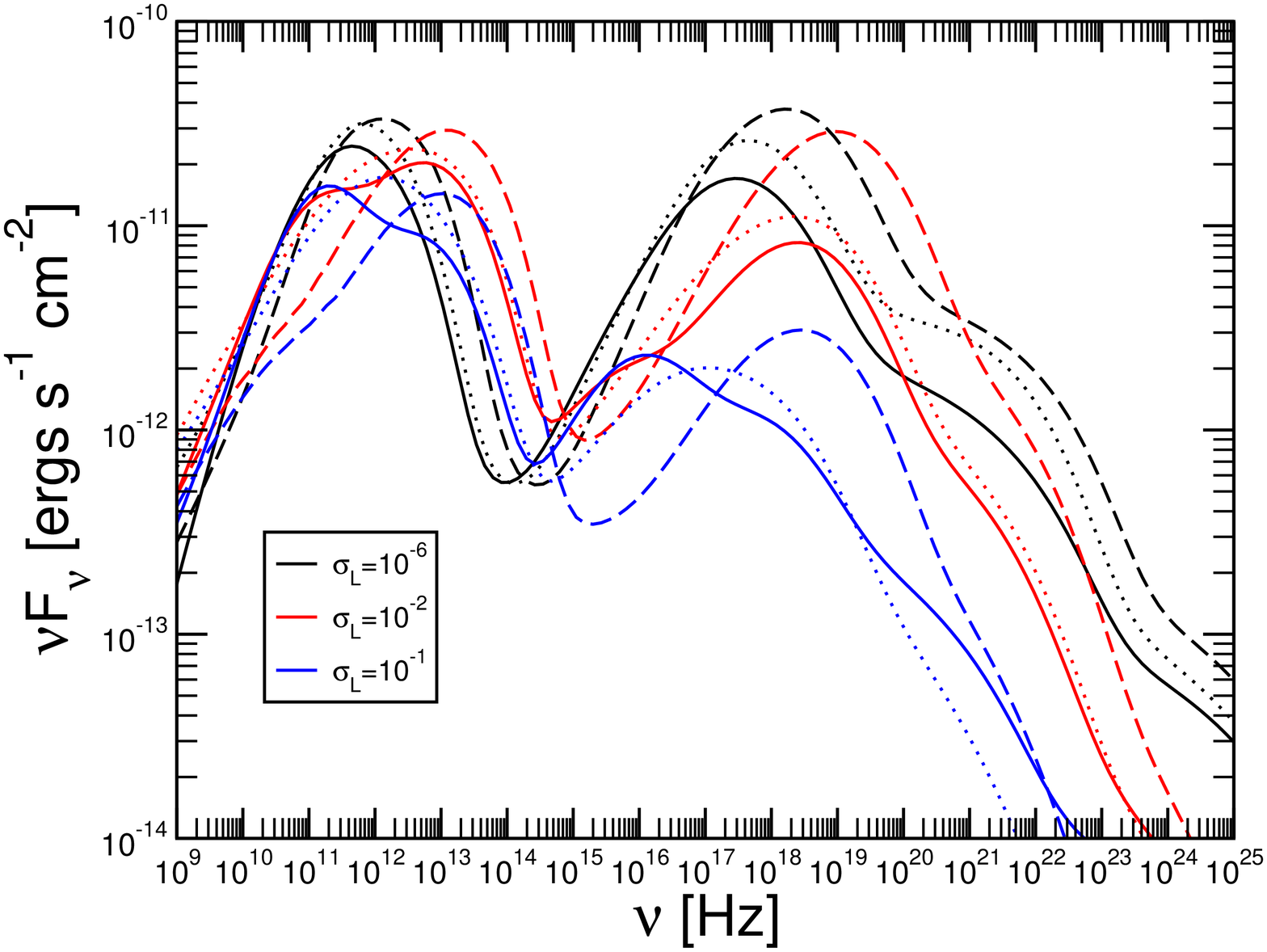}
  \caption{Averaged spectra for different fast shell magnetization,
    $\sigmaL$, with non-thermal particles population fraction
    $\zeta_e = 0.9$ and $0.1$ (left and right panels, respectively). The
    solid, dotted and dashed lines correspond to a magnetization of the
    slower shell $\sigmaR = 10^{-6}, 10^{-2}, 10^{-1}$, respectively. As
    was shown in RMA14, the strongly magnetized fast shells do not display
    a prominent second bump at high frequencies. The synchrotron peak in
    all cases and in both panels, does not surpass $\sim 10^{13}\,$Hz.}
\label{fig:magnetizations}
\end{figure*}

In Fig.~\ref{fig:magnetizations} we show the average spectra produced by
the IS model with different combinations of the faster and slower shells
magnetizations for a fixed EED with $\zetae = 0.9$. In black, red and blue
we represent the models with faster shell magnetization
$\sigmaL = 10^{-6}, 10^{-2} \mbox{ and } 10^{-1}$, respectively. The solid,
dotted and dashed lines correspond to a slower shell magnetization
$\sigmaR = 10^{-6}, 10^{-2} \mbox{ and } 10^{-1}$ respectively. Consistent
with the results in RMA14, the collision of strongly magnetized shells
produces a SSC component dimmer than the synchrotron component. A double
bump outline is reproduced by the model \modelM{10}{1.0}{09}{1} (dashed,
red line) and all the models with $\sigmaL = 10^{-6}$. For most models
$\nu_{\rm syn}$ is situated at $\sim 10^{12}\,$Hz. However, for the cases
with $\sigmaL = 10^{-2}, 10^{-1} \mbox{ and } \sigmaR = 10^{-2}$,
$\nu_{\rm syn}\sim 10^{13}\,$Hz. In both cases, these frequencies reside in
the LSP regime. Remarkably, a change of 4 orders of magnitude in $\sigmaR$
results in an increase of $\lesssim 2$ in the observed flux in models with
an EED dominated by non-thermal electrons ($\zetae=0.9$;
Fig.\,\ref{fig:magnetizations} left panel). In the case of models with a
thermally-dominated EED ($\zetae=0.1$; Fig.\,\ref{fig:magnetizations} right
panel), the change in flux under the same variation of the magnetization of
the slower shell is a bit larger, but still by a factor $\lesssim 6$. In
both cases the larger differences when changing $\sigmaR$ happen in the
decaying side of the spectrum occurring to the right of either the
synchrotron or the SSC peaks. The variation of the magnetization of the
faster shell yields, as expected (MA12; RMA14) larger spectral changes,
especially in the SSC part of the spectrum.

\subsection{Relative Lorentz factor \texorpdfstring{$\Delta g$}{Dg}}
\label{sec:relat-loretz-fact}

\begin{figure*}
  \centering
  \includegraphics[width=8.5cm]{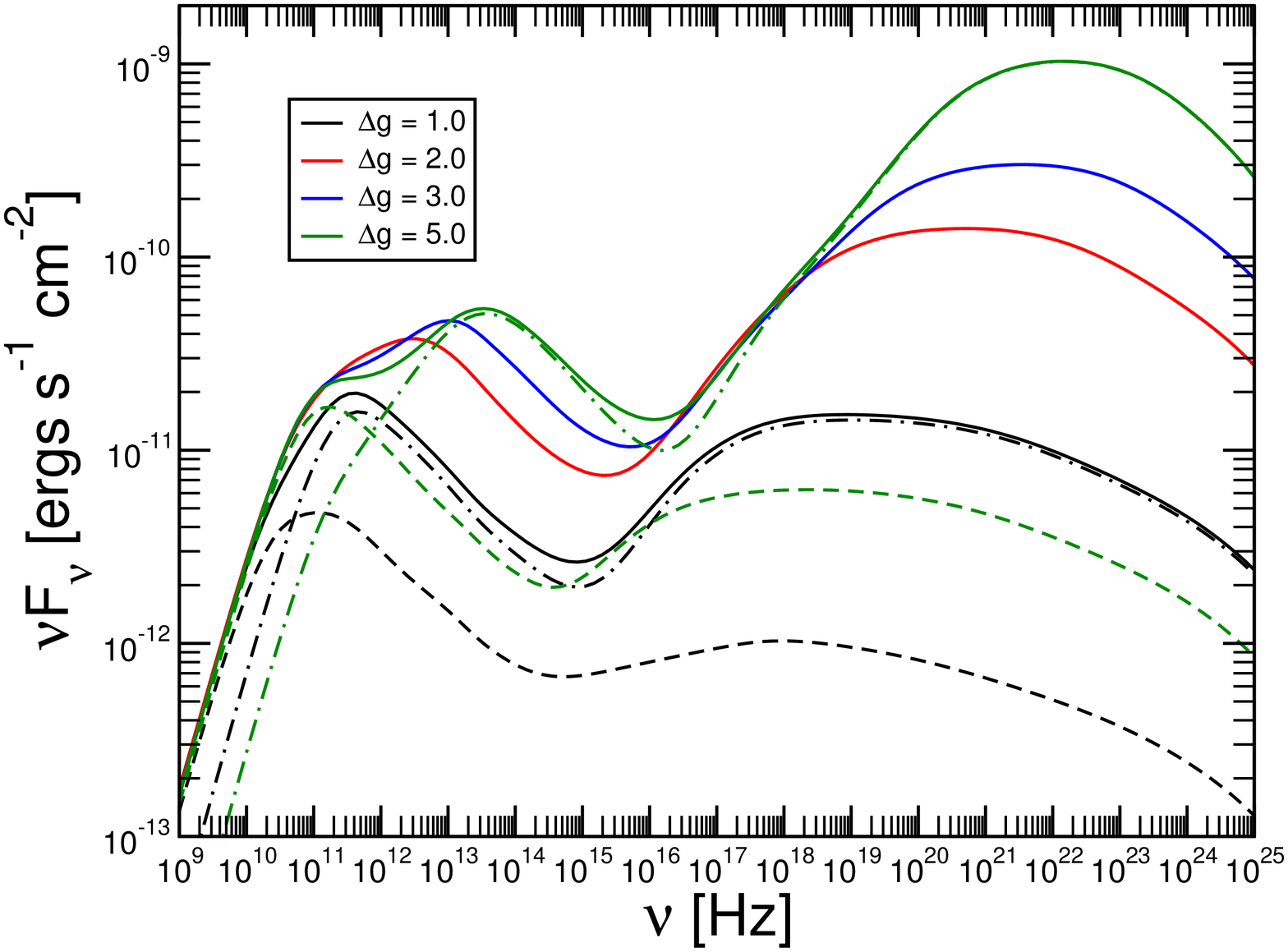}
  \hspace{0.16cm}
  \includegraphics[width=8.5cm]{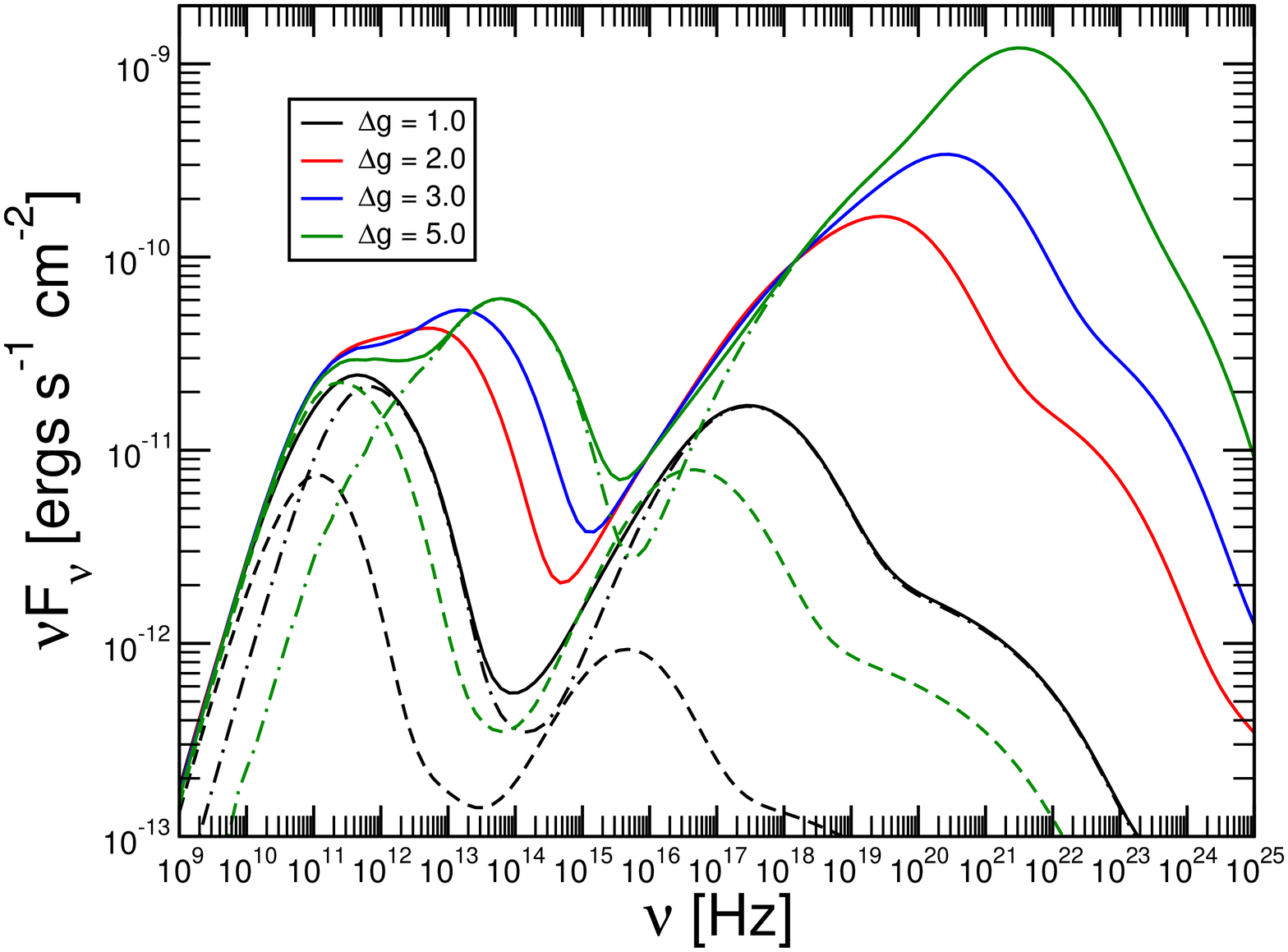}
  \caption{Averaged spectra for different relative Lorentz factors and
    fractions of non-thermal particles. On the left panel we present the
    SED from a particle distribution with $\zetae = 0.9$ while on the right
    panel we show the SED for the same conditions, but with $\zetae =
    0.1$. For the models with $\Delta g=1.0$ (black lines) and
    $\Delta g = 5.0$ (green lines) the FS and the RS individual
    contributions are depicted in dashed and dot-dashed lines,
    respectively. The models depicted are \modelW{10}{(1.0, $\ldots$,
      5.0)}{09}{1} (left panel) and \modelW{10}{(1.0, $\ldots$,
      5.0)}{m1}{1} (right panel).}
\label{fig:Delta-g}
\end{figure*}

In Fig.~\ref{fig:Delta-g} we show the variation of the relative Lorentz
factor, $\Delta g$, for $\zetae = 0.1 \mbox{ and } 0.9$ (\modelW{10}{(1.0,
  $\ldots$, 5.0)}{m1}{1} and \modelW{10}{(1.0, $\ldots$, 5.0)}{09}{1}). The
dashed and dot-dashed lines depict the energy flux coming from the FS and
RS, respectively. The model with $\Delta g = 1.0$ results from the
collision with a fast shell having $\Gamma_{\rm L} = 20$, whereas the case
$\Delta g = 5.0$ assumes that the fast shell moves with
$\Gamma_{\rm L} = 60$ (i.e., slightly above the upper end of the Lorentz
factor distribution for parsec-scale jets; \citealt{Lister:2016AJ}). Both
panels show that the larger the $\Delta g$, the higher the SSC bump. The
colliding shells with relative Lorentz factor $\Delta g = 5.0$ produced a
spectrum with an SSC component one order of magnitude larger than its
synchrotron component. On the other hand, the colliding shells with
relative Lorentz factor $\Delta g = 1.0$ produced a SSC component less
intense than the synchrotron component. Another important feature in these
spectra is the emergence of a second bump in the synchrotron component at
the near infrared ($\sim 10^{14} \unit{Hz}$), which corresponds to emission
coming from the reverse shock. The effect of changing $\zetae$ at high
frequencies is that the larger the non-thermal population of electrons the
broader the SSC component. Moreover, it can be seen that the forward shock
(FS) cannot by itself reproduce the double bump structure of the SED for
blazars, and that the emission coming from the reverse shock (RS) dominates
and clearly shapes the overall spectrum. More specifically, the emission
due to the RS is $\gamma$-ray louder than that of the FS.

The inclusion of a thermal population in the EED combined with a
variation of the relative shell Lorentz factor has a potentially
measurable impact on the blazar spectra modelling. If narrower SSC
peaks and a much steeper decay post-maximum are observed, that could
identify the presence of a dominant thermal emission
(Fig.~\ref{fig:Delta-g}; right). The slope of the $\gamma$-to-TeV
spectrum becomes steeper and more monotonically decaying as $\Delta g$
increases for thermally-dominated EEDs.

\subsection{Lorentz factor of the slower shell}
\label{sec:lorentz-fact-slow}

\begin{figure}
  \centering
  \includegraphics[width=8cm]{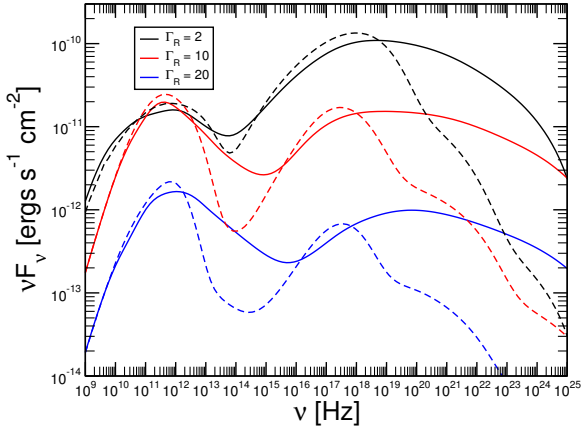}
  \caption{Averaged spectra for weakly magnetized shells with varying
    slower shell bulk Lorentz factor, $\Gamma_{\rm R}$, and two different
    non-thermal particles fractions: $\zetae = 0.9, 0.1$, solid and dashed
    lines respectively.}
\label{fig:lorentz-factor-slow-shell}
\end{figure}

In Fig.~\ref{fig:lorentz-factor-slow-shell} we depict the SEDs resulting
from the collision of weakly magnetized shells with different
$\Gamma_{\rm R}$ and $\zetae$. The solid lines correspond to $\zetae = 0.9$
(models \modelW{(2, 10, 20)}{1.0}{09}{1}) while the dashed lines correspond
to $\zetae = 0.1$ (models\modelW{(2, 10, 20)}{1.0}{m1}{1}). The general
trend is that the brightness of the source suffers an attenuation as
$\Gamma_{\rm R}$ increases, regardless of $\zetae$. From
Eq.~\eqref{eq:numdens} we can see that an increase of the bulk Lorentz
factor of a shell at constant luminosity implies a lower particle density
number. Therefore, less particles are accelerated at the moment of the
collision, which explains the overall flux decrease as $\Gamma_{\rm R}$
increases. Over almost the whole frequency range the brightness of models
depends monotonically on $\Gamma_{\rm R}$, brighter models corresponding to
smaller values of $\Gamma_{\rm R}$. However, the relative importance of the
SSC component does not follow a monotonic dependence. At the lowest value
of $\Gamma_{\rm R}$ the SSC component is brighter than the synchrotron
component by one order of magnitude; with a steeper decay at high
frequencies, though. This monotonic behavior is only broken in the vicinity
of the synchrotron peak when the beaming cone half-opening angle
($\sim 1/\Gamma_{\rm R}$) falls below the angle to the line of sight
($\theta=5^\circ$). This explains the larger synchrotron peak flux when
$\Gamma_{R} = 10$ than when $\Gamma_{R} = 2$. In addition, models with
$\Gamma_{R}=20$ (\modelW{20}{1.0}{(09,m1)}{1}) suffer a greater attenuation
due to Doppler deboosting \citep[see][]{Rueda:2014mn}. In these models the
half-opening angle of the beamed radiation is smaller than the observer
viewing angle, therefore the apparent luminosity decreases.

\subsection{Total luminosity}
\label{sec:total-luminosity}

\begin{figure}
  \centering
  \includegraphics[width=8cm]{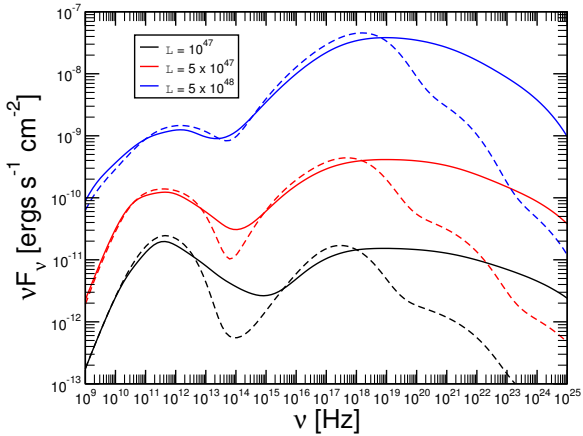}
  \caption{Averaged spectra for different jet total luminosity. Solid and
    dashed lines display the models with $\zetae = 0.9, 0.1$,
    respectively. Different color lines correspond to different values of
    the jet luminosity (see legend).}
\label{fig:luminosity}
\end{figure}

The number of particles accelerated by the internal shocks is an important
quantity in our treatment of EEDs. The number of particles in each shell is
dictated by Eq.~\eqref{eq:numdens}. Such a direct influence of the
luminosity on the number of particles motivates us to study the behaviour
of the SEDs when this parameter is changed. In Fig.~\ref{fig:luminosity} we
show the SEDs produced by the IS model with different total jet
luminosities and values of $\zetae$ (models \modelW{10}{1.0}{(09, m1)}{(1,
  5, 50)}). With solid and dashed lines we differentiate the HDs with
$\zetae = 0.9, 0.1$, respectively, and in black, red and blue the
luminosities $\LL = 10^{47}, 5 \times 10^{47}, 5 \times 10^{48}$,
respectively. The increase in flux of the thermally or non-thermally
dominated cases is rather similar, and follows the expectations. An
increase by 50 in the total luminosity $\LL$ implies an overall increase of
100 in the particle density according to Eq.~\eqref{eq:numdens}. Hence, the
expected increase in flux in the synchrotron component is proportional to
$n_i \sim 100$, while in the SSC component it is proportional to
$n_i^2 \sim 10^4$.

\section{Temperature vs.\ magnetization}
\label{sec:temp-vs-mag}
\begin{figure}
  \centering
  \resizebox{\columnwidth}{!}{
    \input{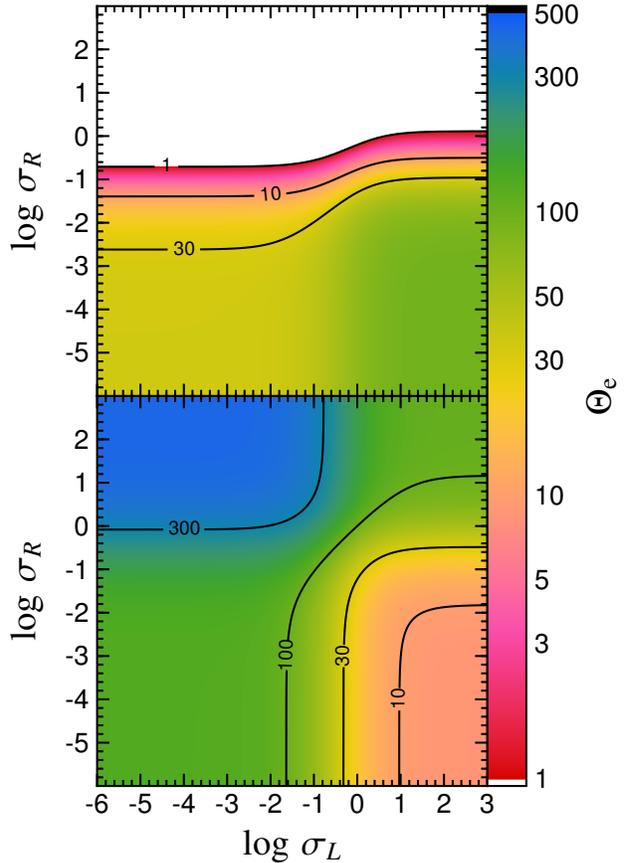}
  }
  \caption{Dependence of the electron temperature on shell
    magnetization. The top and bottom panels show the behaviour of
    $\Thetae$ in the FS and RS, respectively. Contour lines of selected
    temperatures are overlaid in both panels.}
\label{fig:sL-sR-Thetae}
\end{figure}

The fluid temperature $\chi$ is calculated by the exact Riemann solver for
each shell collision. Assuming that the jet is composed of protons and
electrons, the temperature of the electrons in the plasma is
$\Thetae = \chi \mpr / \mel$, where $\mpr$ is the proton mass. In order to
systematically explore the dependence of the temperature on the properties
of the shells we solved a large number of Riemann problems for different
magnetizations and relative Lorentz factor. Here we present the behaviour
of $\Thetae$ in the ISs model in order to obtain insight into the
temperature of the thermal component of the EED in the shocks. In
Fig.~\ref{fig:sL-sR-Thetae} we show the value of $\Thetae$ as a function of
the magnetizations $\sigmaL \mbox{ and } \sigmaR$ for both FS and RS (left
and right panels, respectively).

The hottest region of the RS plane ($\sigmaL<1$ and $\sigmaR>0.1$)
corresponds to the coldest region in the FS plane. Indeed, comparing both
figures we observe that the RS is hotter than the FS wherever
$\sigmaL\lesssim 0.2$ or $\sigmaR>0.1$.  As a result, in most of the
moderately and weakly magnetized models, the radiation produced by the
population of injected electrons that are thermally dominated could come
from the RS. However, for $\sigmaR \lesssim 0.2$ and $\sigmaL \gtrsim 0.2$
the oposite true: the FS is hotter than the RS.%

\begin{figure}
  \centering
  \includegraphics[width=8cm]{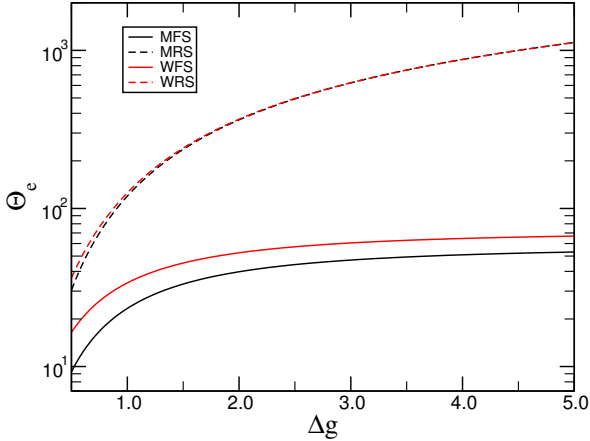}
  \caption{Temperature as a function of the relative Lorentz factor. In
    this figure we show the temperature of both forward (full lines) and
    reverse (dashed lines) shocks for the weakly (red lines) and moderately
    (black lines) magnetized models. The value of the bulk Lorentz factor
    of the slower shell for both magnetization is $\Gamma_{\rm R} = 10$.}
\label{fig:temp-vs-dg}
\end{figure}

In Fig.~\ref{fig:temp-vs-dg} we show the behavior of the electron
temperature $\Thetae$ in terms of the relative Lorentz factor $\Delta g$
between the colliding shells for the FS and RS. In accordance with
figures~\ref{fig:sL-sR-Thetae}, the reverse shock is hotter than the
forward shock. As the relative Lorentz factor $\Delta g$ grows the
temperature of the reverse shock tends to grow while the forward shock
seems to be approaching asymptotically to a value, which depends slightly
on the magnetization (the larger the magnetization the smaller the
asymptotic temperature). Values $\Delta g > 5$ are inconsistent with the
blazar scenario, for a fixed value $\Gamma_{\rm R}=10$, since they would
imply that the faster shell was moving at $\Gamma_{\rm L} > 60$ (in excess
of the maximum values of the Lorentz factor for the bulk motion inferred
for blazars).

From figures~\ref{fig:sL-sR-Thetae} and~\ref{fig:temp-vs-dg} we can infer
that $\Thetae$ does not only depend on the velocity of the fluid but also
on its magnetization. Therefore, we conclude that this degeneracy makes the
determination of $\Theta_e$ a very difficult task.

\section{Discussion and conclusions}
\label{sec:discuss}

In this work we introduce a hybrid thermal-non-thermal electron
distribution into the internal shock model for blazars. To account for
the fact that the thermal component of the HD extends to very low
electron Lorentz factors, we also introduce a cyclo-synchrotron code
that enables us to compute the non-thermal emission from electrons
with arbitrary Lorentz factor. We show that our method for treating
the temporal evolution of the HD and the calculation of MBS emission
can be performed efficiently and with sufficient accuracy. The method
is implemented as a generalization of the numerical code of MA12.

To test the influence of the fraction of non-thermal particles $\zetae$ in
the overall HD we apply the new method to the case of a blazar with
$\LL = 10^{47} \unit{erg} \unit{s}^{-1}$
(Fig.~\ref{fig:nonthermal-pop}). Considering only MBS and SSC emission
processes we see that increasing $\zetae$ (i.e., the distribution becoming
more non-thermal) has as a consequence a shallower valley between the two
spectral peaks, while the SSC emission extends to higher energies. In other
words, a HD of mostly thermal particles emits only up to MeV (except when
$\Delta g \sim 5$; see Fig.~\ref{fig:Delta-g}). This would mean that the
emission in the GeV range for the thermally-dominated HD cannot come from
the SSC and would have to be produced by the EIC (not considered
here). Furthermore, Fig.~\ref{fig:nonthermal-pop} confirms that also for
low $\zetae$ highly-magnetized blazar jets seem to be observationally
excluded because their SSC peak is too dim.

Another effect of decreasing $\zetae$ is the shift of the SSC peak to lower
frequencies and the narrowing of the high-frequency spectral bump, while at
the same time the synchrotron peak and flux do not change appreciably. This
means that (excluding possible effects from varying EIC) the Compton
dominance (ratio of internal Compton and cyclosynchrotron luminosity) can
be changed by varying $\zetae$, while the peak MBS frequency remains
constant. In other words, for all other parameters remaining constant, the
variations in $\zetae$ may explain the vertical scatter in the distribution
of FSRQs and BL Lacs in the peak synchrotron frequency- Compton dominance
parameter space \citep[see e.g., Fig. 5 in][]{Finke:2013aj}. Changing
$\zetae$ appears to not be able to change the blazar class.

Regarding the variations of the shell magnetization
(Sec.~\ref{sec:magnetization}), relative Lorentz factor
(Sec.~\ref{sec:relat-loretz-fact}) and the bulk Lorentz factor
(Sec.~\ref{sec:lorentz-fact-slow}), the results are consistent with those
of RMA14. In this work we performed a more detailed study of the influence
of the magnetization than in the previous paper since now we study $9$
possible combinations of faster and slower shell magnetizations, instead of
only three in RMA14. The truly novel result of this work is that the RMA14
trend generally holds for the thermally-dominated HD as well (right panel
in Fig.~\ref{fig:magnetizations}), with the difference that the collision
of $(\sigmaL = 0.1, \sigma_{\rm R} = 0.1)$ shells produces a double-peaked
spectrum for $\zeta_e=0.1$, while its non-thermally dominated equivalent
does not (blue dashed lines in Fig.~\ref{fig:magnetizations}). Even so, the
SSC component remains very dim for very magnetized shells.

Regarding $\Delta g$, the RS emission (dot-dashed lines in
Fig.~\ref{fig:Delta-g}) is crucial for reproducing the blazar
spectrum. Therefore, in the case of $\zeta_e\ll 1$ the temperature of the
RS is one of the most important parameters. Since this temperature
increases with $\Delta g$ (Fig.~\ref{fig:temp-vs-dg}), the effect of
$\Delta g$ on the MBS and the SSC peak frequencies and fluxes is
qualitatively similar to that of the non-thermal electron distribution
(Fig.~\ref{fig:Delta-g}; see also RMA14). The changes induced by variations
of $\Gamma_{\rm R}$ (Fig.~\ref{fig:lorentz-factor-slow-shell}) are
independent of the thermal/non-thermal EED content and agree with
RMA14. The effects of the increase in total jet luminosity are visible both
for $\zetae = 0.1$ and $\zetae = 0.9$. Varying the luminosity by a factor
$50$ increases the MBS flux by $\sim 10^2$ and the SSC flux by $\sim
10^4$. The relation between spectral components is very similar to the
variations of $\Gamma_{\rm R}$, i.e. the increase in ${\cal L}$ is similar
to a decrease in $\Gamma_{\rm R}$.

Overall, we show that the inclusion of the full cyclo-synchrotron
treatment, motivated by the significant low-energy component of the HD, has
a moderate effect on the blazar spectrum at optical-to-$\gamma$ ray
frequencies. However, at lower frequencies (e.g., below 1\,GHz) where the
self-absorption may play a role the differences between the synchrotron and
the MBS will be more severe. We plan to include the effect of absorption in
a future work as well as the effects by EIC emission.

\section*{Acknowledgements}

We acknowledge the support from the European Research Council (grant
CAMAP-259276), and the partial support of grants AYA2015-66899-C2-1-P,
AYA2013-40979-P and PROMETEO-II-2014-069. We also thank to the Mexican
Council for Science and Technology (CONACYT) for the financial support
with a PhD grand for studies abroad. Part of the computations were
performed in the facilities of the Spanish Supercomputing Network on
the clusters \textit{lluisvives} and \textit{Tirant}.

\appendix

\section{The RMA function}
\label{sec:rma-function}

\begin{figure*}
  \centering
  \includegraphics[width=8.5cm]{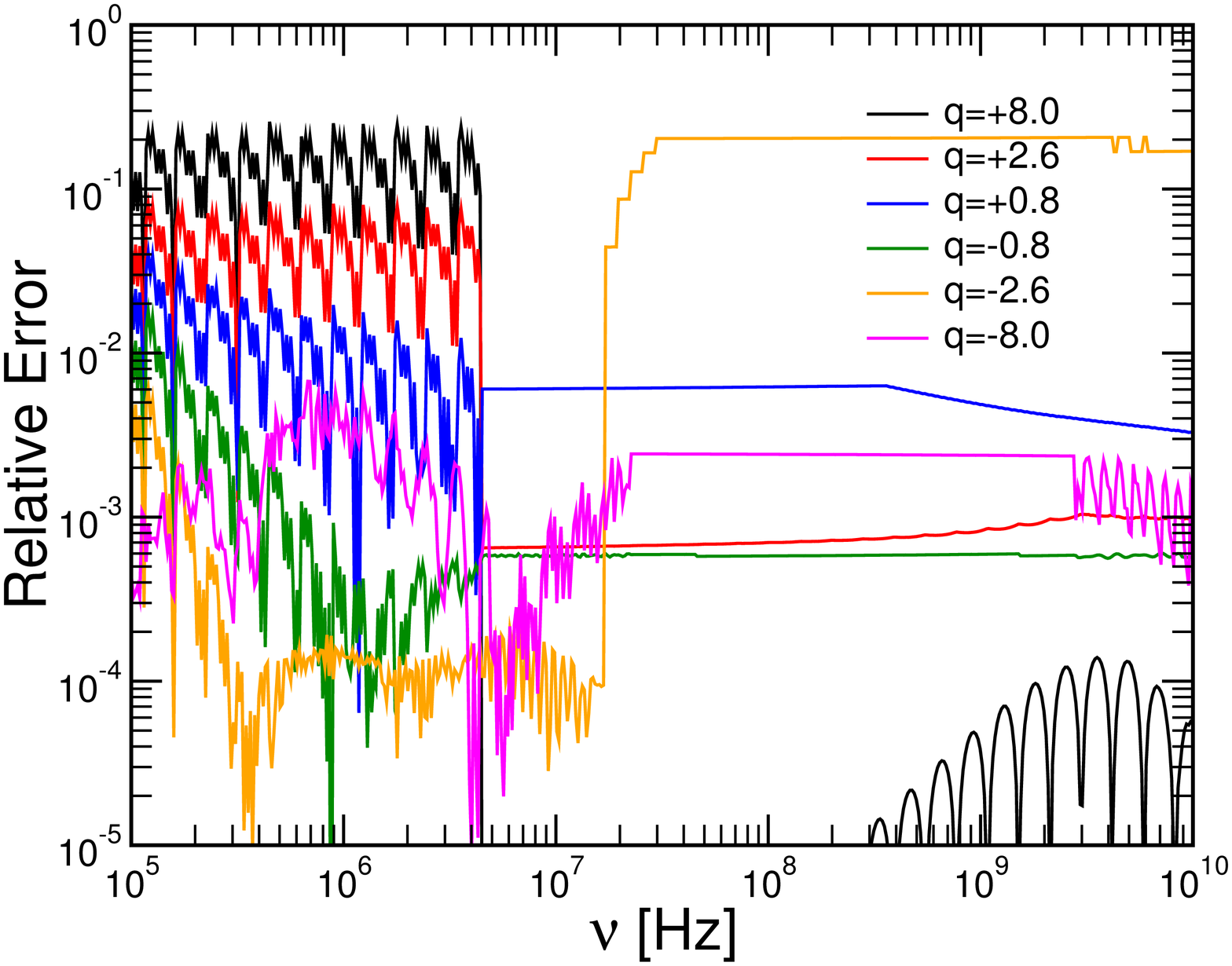}
  \ \hspace{0.25cm}
  \includegraphics[width=8.5cm]{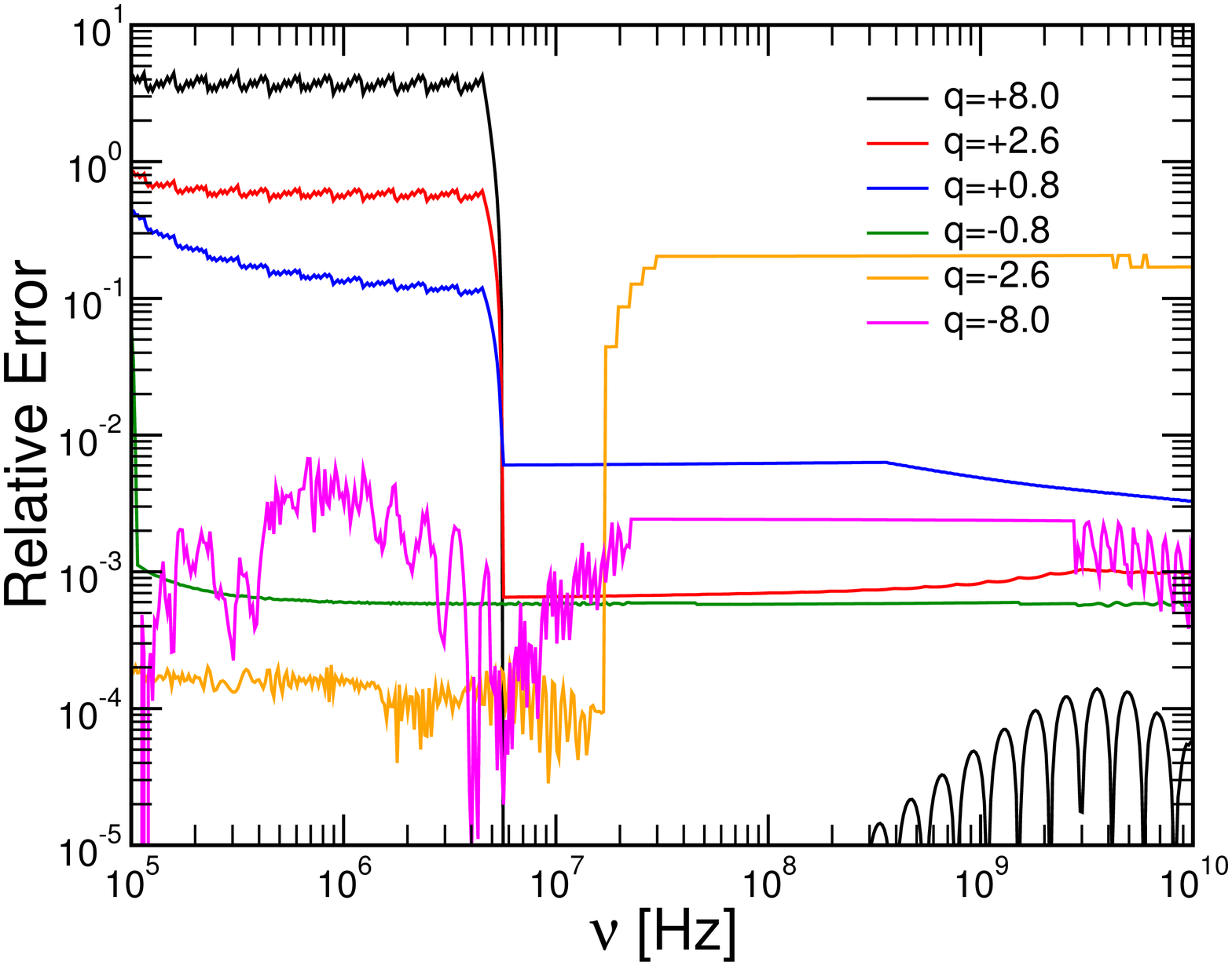}
  \caption{The relative error between emissivity for a power-law
    distribution of electrons computed from the MBS interpolation tables
    and performing numerical integration of the RMA function. Each of the
    different colours represent cases with different power-law indices,
    $q$, of the non-thermal EED. In the left and right panels we show the
    relative error considering $a = 0.8$ and $a=1$ in
    Eq.~\ref{eq:rma-func-a}, respectively.}
\label{fig:rel-err-jnu}
\end{figure*}

The formula for the pitch-angle averaged synchrotron power of a single
ultrarelativistic electron was derived in e.g., \cite{Crusius:1986vu} and
afterwards, an accurate approximation of it was discovered by
\citet{Schlickeiser:2007hp}. Both expressions assume a continuum spectrum
for all $\gamma$, so that they cannot be applied directly to the
calculation of the discrete low-frequency, low-$\gamma$ cyclotron
emission. In particular, these formulae do not take into account the fact
that for slow electrons there is no emission below the gyrofrequency
$\nu_{b}$. Nevertheless, the expression in \citet{Schlickeiser:2007hp} is
analytic, which makes it very convenient for a fast numerical
implementation. We use the Eq.~(16) in \citet{Schlickeiser:2007hp} to
define the function\footnote{In the process of looking for a good
  analytical approximation to CS[x] we tried to generalize the approach
  made by SL07 by fitting the numerical data with
  $\widetilde{CS}[x; a, b, c] := x^{-a} / (b + x^{c} e^{x})$. We found,
  nevertheless, that the quality of the approximation of SL07 to CS86 was,
  indeed, good enough for our purposes. However, it has been shown by
  \citet{Finke:2008cz} that a piece-wise approach may lead to better
  fits. As a future work, we will try to improve the $RMA$ function testing
  the piece-wise approach of~\cite{Finke:2008cz}.}
\begin{equation} \label{eq:rma-func-a}
RMA[x] :=
\begin{cases}
  x\, CS[x] & x > 2 a / (3 \gamma^3) \\
  0         & \text{otherwise}
\end{cases}
,
\end{equation}
where $a$ is a numerical constant. The location of the cut-off; i.e., the
value of $a$ in Eq.~\eqref{eq:rma-func-a}, is very important. In
Fig.~\ref{fig:rel-err-jnu} we show the relative error of the emissivity
using $RMA[x]$ compared to full MBS treatment. We assume a pure power-law
distribution of electrons with different power-law indices and use two
different values of the cut-off constant: $a = 0.8$ and $a=1$. The magnetic
field for this test was $B = 10 \unit{G}$ and the minimum and maximum
Lorentz factors $\gminnth = 5, \gmaxnth = 500$, respectively. At low
frequencies the errors are large because there the emission is dominated by
harmonics and is thus not well represented by a continuous $RMA[x]$
function. Nevertheless, choosing an appropriate value for $a$ can decrease
the errors in that region from $\sim 350$\% ($a = 1$, right panel) to
$\sim 25$\% ($a = 0.8$, left panel). The relative error of the cases with
power-law indices $q < 0$ are always below 1, and is somewhat lower for
$a = 1$ than for $a = 0.8$. However, since we want the relative error to be
the lowest for all power-law indices, we choose the cut-off constant
$a = 0.8$\footnote{Further scanning of the values of $a$ showed that a
  decrement of this parameter rises the relative error at low
  frequencies.}.
\section{The \texorpdfstring{$\nchi^{2} I_{1}$}{X_2 I_1} interpolation table}
\label{sec:interpol-table}
The interpolation table of
$\tilde{I}_{1}(\gamma, \nchi) := \nchi^{2} I_{1}(\nchi, \gamma)$ was built
integrating Eq.~\eqref{eq:I1} using the Gauss-Legendre quadrature with 120
nodal points for values of $\gamma \leq \gamma_{\rm up}$ and
$\nchi \le 10^{2}$. For $\nchi > 10^{2}$ and $\gamma > \gamma_{\rm up}$ we
employ the approximate expression
$\tilde{I}_{1}(\gamma, \nchi) \simeq RMA[x]$ (see
Eq.\,\eqref{eq:RMA-I1}). The numerical calculations of the Bessel functions
were performed using the tool \texttt{my\_Bessel\_J} developed in
\citet{Leung:2011aj}. Computing $\tilde{I}_{1}(\gamma, \nchi)$ for $\gamma$
and $\nchi$ outside this region is computationally
challenging. Fortunately, in the ultrarelativistic regime we can
approximate $\tilde{I}_{1}(\gamma, \nchi)$ using the $RMA$ function (see
App.~\ref{sec:rma-function}). In the $\gamma$ direction $\tilde{I}_{1}$ is
approximated using Chebyshev interpolation (for each $\nchi$ separately).
\begin{figure}
  \centering
  \resizebox{\columnwidth}{!}{
    \input{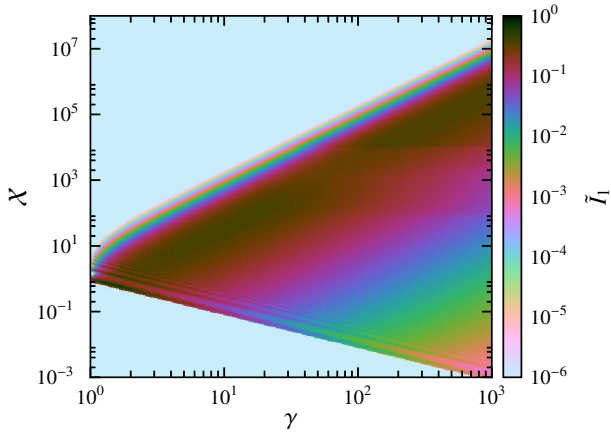}
  }
  \caption{$\nchi^2 I_1$ as a function of $\nchi$ and $\gamma$. The
    emission is zero in the light blue region. We also note that for
    arbitrary $\gamma$ there is a sufficiently low $\nchi$ so that the
    emission is in the form of harmonics.}
\label{fig:I1-chi-gamma}
\end{figure}

Special care has to be devoted to the zero emission regions below
$\nchi_{1}(\gamma)$ and above $\nchi = 100$ (light blue triangular
zones in Fig.~\ref{fig:I1-chi-gamma}), since including those regions
can cause a bad numerical behaviour of Chebyshev interpolation. In
order to avoid this, we constructed a Lorentz factors array
$\{\hat{\gamma}_{\min}(\nchi)\}$ containing the minimum Lorentz factor
above which the emission is non-negligible for every value of $\nchi$;
i.e., $\{\hat{\gamma}_{\min}(\nchi)\}$ is a set of lower interval
limits for the Chebyshev interpolation (instead of $\gamma = 1$).
\begin{figure}
  \centering
  \includegraphics[width=8cm]{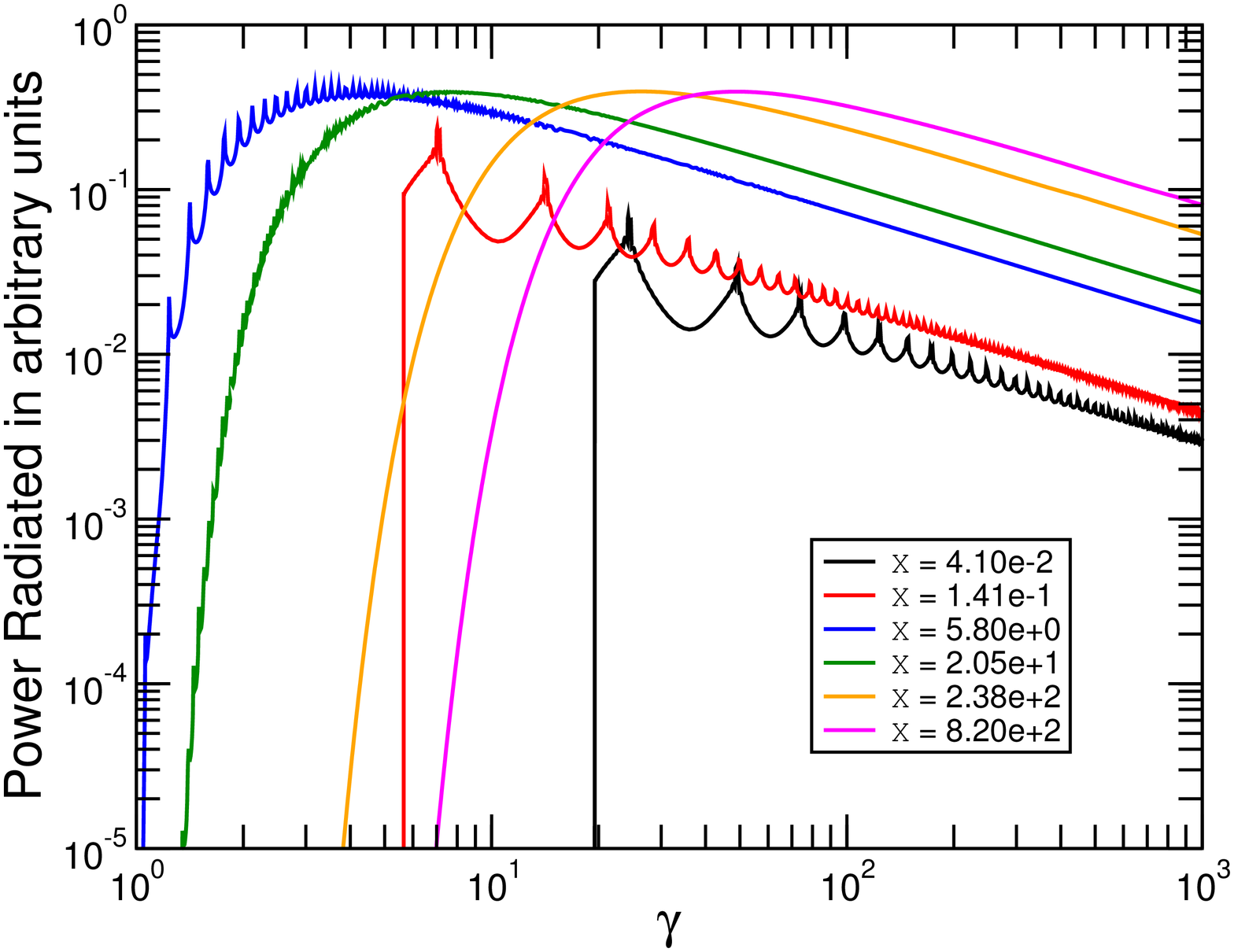}
  \caption{Similar to Fig.~\ref{fig:chamba-sl-rma} but for a fixed
    $\nchi$. The black and red lines depict the radiated power for
    $\nchi < \nchi_{1}$. The break at low $\gamma$ is set by hand
    considering the cut-off criteria described in
    Sec.~\ref{sec:rma-function}. The blue and green lines correspond to
    $\nchi_{1} \leq \nchi < 100$. The orange and magenta lines correspond
    to $\nchi \geq 100$.}
\label{fig:jnuvsgamma-newtable}
\end{figure}

\subsection{Minimum Lorentz factors for \texorpdfstring{$\nchi < \nchi_{1}$}{X < X1}}
\label{sec:chi-less-chi1}

Numerical calculations of the cyclo-synchrotron radiated power show that
the frequency of the first harmonic behaves as
$\nchi_{1}(\gamma) = 1 / \gamma$. In App.~\ref{sec:rma-function} we show
the cut-off criterion chosen to include as much power radiated as possible
while avoiding the zero emission frequencies below $\nchi_{1}(\gamma)$. We
follow a similar procedure to construct the array
$\{\hat{\gamma}_{\min}(\nchi)\}$; i.e.,
$\hat{\gamma}_{\min}(\nchi) = 0.8 / \nchi$.

\subsection{Minimum Lorentz factors for \texorpdfstring{$\nchi \geq 100$}{X
  >= 100}}
\label{sec:chi-geq-1e2}

Finding $\hat{\gamma}_{\min}(\nchi)$ for this side of the
spectrum requires of a two-step procedure:
\begin{enumerate}
\item For every $\nchi$ the bisection method was employed to find the value
  of $\gamma$ at which $\tilde{I}_{1}$ is well below its maximum value.
\item A linear fit (in logarithmic space) was performed with the values of
  $\gamma$ found in the previous step.
\end{enumerate}

We used the formula obtained from the fit to estimate the values of
$\hat{\gamma}_{\min}(\nchi)$ in this region.

\subsection{Minimum Lorentz factors for \texorpdfstring{$\nchi_{1} \leq
    \nchi < 100$}{X_1 <= 100}}
\label{sec:chi1-leq-chi-less-1e2}

Our calculations showed that in the region where $1 \leq \nchi < 100$ there
is practically no zero radiation region in the $\gamma$ direction (see
Fig.~\ref{fig:I1-chi-gamma}). Since this region is above the first harmonic
$\nchi_{1}$, neither the criterion used in App.~\ref{sec:chi-less-chi1} nor
the bisection procedure employed in App.~\ref{sec:chi-geq-1e2} can be used
here since the profile of $\tilde{I}_{1}$ is too steep at $\gamma \sim 1$
(see Fig.~\ref{fig:jnuvsgamma-newtable}). Applying a bisection method leads
to an oscillating $\hat{\gamma}_{\min}(\nchi)$ which, produces numerical
problems when interpolating from the table. We verified that a constant
Lorentz factor minimum threshold close to 1 produces good results in this
region. Thus, we employ the input parameter $\gminth$ for this
purpose. Normally we use the numerical value $\gminth \approx 1.005037815$
which corresponds to the Lorentz factor of a particle with $\beta=0.1$. The
exact value $\gamma = 1$ cannot be used as threshold because it corresponds
to $\beta=0$, causing problems in e.g., the resonance condition
(Eq.~\eqref{eq:reson-cond}) and the subsequent equations.

\subsection{Calculation of
  \texorpdfstring{$\nchi^{2} I_{1}(\nchi, \gamma)$}{X2 I1(X, g)} using the
  interpolation table}
\label{sec:reconstr-radi-power}

\begin{figure}
  \centering
  \resizebox{\columnwidth}{!}{
    \input{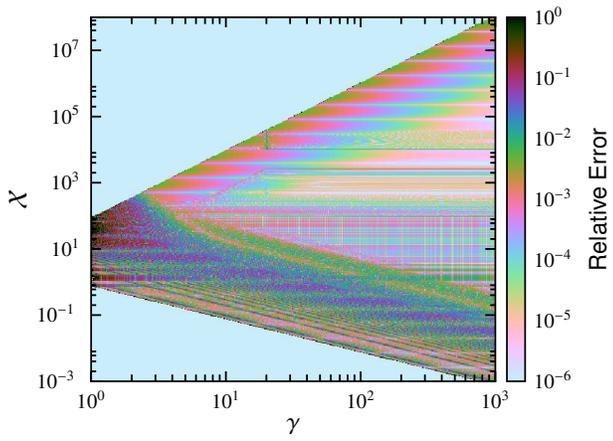}
  }
  \caption{Similar to Fig.~\ref{fig:I1-chi-gamma}, but showing the
    relative error between the data obtained using numerical
    integration and the values interpolated from the table. The
    resolution of the plot is $1024 \times 1024$ points.}
\label{fig:Rel-Err-grad}
\end{figure}

The usage of $\tilde{I}_{1}$ requires a two-step procedure: (1) Chebyshev
interpolation from the Chebyshev coefficients in the $\gamma$ direction and
(2) a linear interpolation in the $\nchi$ direction using the values
obtained in the first step. The accuracy of the reconstruction routine can
be seen in Fig.~\ref{fig:Rel-Err-grad}. The test was performed on a grid of
$1024 \times 1024$. The relative error in most of the points is
$\lesssim 1$\%.

\bibliographystyle{mn2e}
\bibliography{HybDis}

\end{document}

%% file: Figures/Fig02.tex
\begin{tikzpicture}[node font=\large, scale=1.5]
  \coordinate (O) at (0,0);

  \fill[color=gray!40!white] (0,0) rectangle (3.85,3.85);

  \fill[color=orange!40!white] (0.3,0.0) rectangle (0.35,3.85);
  \fill[color=cyan!40!white] (0.35,0.4) rectangle (3.3,3.2);

  \fill[color=red!60!black] (0,0) rectangle (0.3,3.85);
  \node[rotate=90, color=white] at (0.15,2) {$\gamma < 1$};

  \draw[thick, arrows={Latex[]-Latex[]}] (0,4) node (yaxis) [left] {$\mathcal{X}$} |- (4,0)
  node (xaxis) [below] {$\xi$};

  \draw[dashed, thick] (0,3.2) node[left] {$\mathcal{X}_{\max}$} -- (3.85,3.2);
  \draw[dashed, thick] (0,0.4) node[left] {$\mathcal{X}_{\min}$} -- (3.85,0.4);
  \draw[dashed, thick] (0.35,0) node[below] {$\xi_{\min}$} -- (0.35,3.85);
  \draw[dashed, thick] (3.3,0) node[below] {$1$} -- (3.3,3.85);

  \node at (1.8,3.5) {\texttt{uinterp}};
  \node at (1.8,2) {\texttt{disTable}};


\end{tikzpicture}
